\documentclass[]{aastex63}

\received{07 October 2020}
\revised{}
\accepted{30 October 2020}
\submitjournal{AJ}

\shorttitle{Oxygen Abundance in the Solar neighbourhood}
\shortauthors{Franchini et al.}
\graphicspath{{./}{figures/}}

\begin{document}

\title{The Gaia-ESO Survey: Oxygen abundance in the Galactic thin and thick disks \footnote{Based on observations collected with the FLAMES instrument at VLT/UT2 telescope (Paranal Observatory, ESO, Chile), for the Gaia--ESO Large Public Spectroscopic Survey (188.B--3002, 193.B--0936).}}

\correspondingauthor{Mariagrazia Franchini}
\email{Mariagrazia.franchini@inaf.it}

\author[0000-0001-5611-2333]{Mariagrazia Franchini}
\affil{INAF - Osservatorio Astronomico di Trieste, Via G. B. Tiepolo 11, Trieste, I-34143, Italy }

\author[0000-0002-3319-6375]{Carlo Morossi}
\affiliation{INAF - Osservatorio Astronomico di Trieste, Via G. B. Tiepolo 11, Trieste, I-34143, Italy }

\author[0000-0003-3168-2289]{Paolo Di Marcantonio}
\affiliation{INAF - Osservatorio Astronomico di Trieste, Via G. B. Tiepolo 11, Trieste, I-34143, Italy }

\author[0000-0003-0407-8115]{Miguel Chavez}
\affiliation{Instituto Nacional de Astrof\'isica, \'Optica y Electr\'onica, Luis Enrique Erro 1,  72840 Tonantzintla, Puebla, Mexico}

\author[0000-0002-0601-6199]{Vardan  Adibekyan}
\affiliation{Instituto de Astrof\'isica e Ci\^encias do Espa\c{c}o, Universidade do Porto, CAUP, Rua das Estrelas, 4150-762 Porto, Portugal  }
\affiliation{Departamento de F\'{\i}sica e Astronomia, Faculdade de Ci\^encias, Universidade do Porto, Rua do Campo Alegre, 4169-007 Porto, Portugal}

\author[0000-0003-3978-1409]{Thomas Bensby}
\affiliation{Lund Observatory, Department of Astronomy and Theoretical Physics, Box 43, SE-221 00 Lund, Sweden}

\author[0000-0002-0338-7883]{Angela Bragaglia}
\affiliation{INAF - Osservatorio di Astrofisica e Scienza dello Spazio di Bologna, Via Gobetti 93/3 I-40129 Bologna, Italy}

\author[0000-0001-9091-5666]{Anais Gonneau}
\affiliation{Institute of Astronomy University of Cambridge
Madingley Road Cambridge CB3 0HA (UK) }

\author[0000-0001-6825-1066]{Ulrike Heiter}   
\affiliation{Observational Astrophysics, Department of Physics and Astronomy, Uppsala University, Box 516, 75120 Uppsala, Sweden}

\author[0000-0002-9035-3920]{Georges Kordopatis}
\affiliation{Universit\'e C\^ote d'Azur, Observatoire de la C\^ote d'Azur, Laboratoire Lagrange, CNRS UMR 7293
CS 34229 06304 Nice Cedex 04 - France}

\author[0000-0003-4486-6802]{Laura Magrini} 
\affiliation{INAF - Osservatorio Astrofisico di Arcetri, Largo E. Fermi 5, Florence  I-50125,  Italy}

\author[0000-0002-0845-6171]{Donatella Romano}
\affiliation{INAF - Osservatorio di Astrofisica e Scienza dello Spazio di Bologna, Via Gobetti 93/3 I-40129 Bologna, Italy}

\author[0000-0002-2285-8708]{Luca Sbordone} 
\affiliation{European Southern Observatory, Alonso de Cordova 3107 Vitacura, Santiago de Chile, Chile}

\author[0000-0003-0942-7855]{Rodolfo Smiljanic}
\affiliation{Nicolaus Copernicus Astronomical Center, Polish Academy of Sciences, ul. Bartycka 18, 00-716, Warsaw, Poland}

\author[0000-0001-7672-154X]{Gra{\v z}ina Tautvai{\v s}ien{\. e}}
\affiliation{Institute of Theoretical Physics and Astronomy, Vilnius University, Sauletekio av. 3, 10258, Vilnius Lithuania}

\author[0000-0003-4632-0213]{Gerry Gilmore}
\affiliation{Institute of Astronomy, University of Cambridge, Madingley Road, Cambridge CB3 0HA, United Kingdom}

\author[0000-0003-2438-0899]{Sofia Randich}
\affiliation{INAF - Osservatorio Astrofisico di Arcetri, Largo E. Fermi 5, 50125, Florence, Italy}

\author[0000-0001-7868-7031]{Amelia Bayo}
\affiliation{Instituto de F\'isica y Astronom\'ia, Universidad de Valpara\'iso, Avda. Gran Breta\~na 1111, Valpara\'iso, Chile}
\affiliation{N\'ucleo Milenio de Formaci\'on Planetaria, NPF, Universidad de Valpara\'iso, Chile}

\author[0000-0002-0155-9434]{Giovanni Carraro}   
\affiliation{Dipartimento di Fisica e Astronomia, Universit\`a di Padova, Vicolo dell'Osservatorio 3,  Padova,   I-35122, Italy}

\author[0000-0003-0304-9910]{Lorenzo Morbidelli} 
\affiliation{INAF - Osservatorio Astrofisico di Arcetri, Largo E. Fermi 5, Florence  I-50125,  Italy}

\author[0000-0001-6081-379X]{Simone Zaggia}   
\affiliation{INAF - Osservatorio Astronomico di Padova, Vicolo dell'Osservatorio 5, Padova,  I-35122, Italy }




\begin{abstract}

We analyze the oxygen abundances of a stellar sample representative of the two major Galactic populations: the thin and thick disks. The aim is to investigate the differences between members of the Galactic disks and to contribute to the understanding on the origin of oxygen chemical enrichment in the Galaxy. The analysis is based on the [O\,{\sc i}]=6300.30\,\AA~ oxygen line in HR spectra ($R\sim$52,500) obtained from the GES Survey. By comparing the observed spectra with a theoretical dataset, computed in LTE with the SPECTRUM synthesis and ATLAS12 codes, we derive the oxygen abundances of 516 FGK dwarfs for which we have previously measured carbon abundances. Based on kinematic, chemical and dynamical considerations we identify 20 thin and 365 thick disk members. We study potential trends of both subsamples in terms of their chemistry ([O/H],  [O/Fe], [O/Mg], and [C/O] versus [Fe/H] and [Mg/H]), age, and position in the Galaxy. Main results are: (a) [O/H] and [O/Fe] ratios versus [Fe/H] show systematic differences between thin  and thick disk stars with enhanced O abundance of  thick disk stars with respect to thin disk members and a monotonic decrement of [O/Fe] with increasing metallicity, even at metal-rich regime;  (b) a smooth correlation of  [O/Mg] with age in both populations, suggesting that this abundance ratio can be a good proxy of stellar ages within the Milky Way; (c) thin disk members with [Fe/H]$\simeq0$ display a [C/O] ratio smaller than the solar value,  suggesting a possibly outward migration of the Sun from lower Galactocentric radii.

\end{abstract}

\keywords{Late-type stars -- Stellar abundances -- Stellar ages -- Galactic stellar disks}


\section{Introduction}
\label{sec:intro}

Oxygen is the third most abundant element in the universe and is produced by hydro-static burning in massive stars and then essentially mostly dispersed by the Type II supernovae \citep{WOO95}. Oxygen and its isotopes provide fundamental evolutionary information in a number of astrophysical topics from exoplanets to galaxies, including, of particular importance, studies on the chemical evolution of the Milky Way \citep{MAT01}. 
Moreover, in recent years, numerous studies have been devoted to investigate whether or
not there is a correlation between the stellar oxygen abundance and the presence of exo-planets \citep[see][]{BRE16, NIS18, PAV19} 

Considerable efforts have been made in the literature in recent years  in deriving accurate O abundance in  late type stars. Unfortunately, the   atomic lines suitable for deriving O abundance are not numerous in the visual spectrum of FGK stars and all of them present some difficult issues.  The most commonly oxygen lines used in literature are the  7773\,\AA~ triplet lines (OI=7772, 7774, and  7775\,\AA), the high-excitation OI 6158\,\AA~line, and the forbidden lines [O\,{\sc i}] 6300\,\AA~and 6363\,\AA.
The OI triplet lines are relatively easily accessible  and located in a spectral region not affected by blending lines, but they are particularly sensitive to  3D non-local thermodynamic equilibrium (NLTE) \citep[e.g.][]{ASP09, AMA15, AMA16, AMA19}. 
On the other hand, the other lines,  less problematic for NLTE effects, are quite weak and, therefore, difficult to measure at relatively low and high temperatures, respectively. An accurate analysis and study of these lines was performed by \citet{BER15} who also discuss the influence of the quality and of the signal-to-noise ratio (SNR)  of the spectra on the discrepancies of results obtained according to the different lines used. Moreover, some of these lines can be  severely contaminated by the presence of telluric lines. 

A different approach uses the strengths of the large number of molecular lines of OH in the near-UV and near-IR,  \citep{ISR98,ISR01a,ISR01b,BOE99a,BOE99b,MEL01} as indicator of oxygen abundance. However, the OH lines, and thus the oxygen abundance derived from them, are very sensitive to the adopted effective
temperature \citep{JON18} and to 3D granulation effects \citep{GON10}. 

 Since different oxygen diagnostics  provide  discordant results \citep[see for example][]{BER15}, the distribution of the [O/Fe] abundance ratio in stars across the Galactic disk(s) is still under debate. Many papers in literature are focused on the understanding of the discrepancies derived by the different lines adopted to estimate oxygen abundance and on their effect on  the analysis of oxygen trends in the disk populations of our Galaxy \citep[e.g.][]{BEN04,BER15}. 

Nowadays, in addition to all these numerous and accurate works, we have also the huge amount of data provided by large modern spectroscopic surveys, such as the Gaia-ESO public spectroscopic Survey \citep[GES,][ESO programmes 188.B-3002
and 193.B-0936]{GIL12, RAN13}, the Apache Point Observatory Galactic Evolution Experiment (APOGEE, \citealt{MAJ17}), and the GALactic Archaeology with HERMES (GALAH, \citealt{DES15}), that, providing multi-elemental abundance ratios of thousands of stars belonging to different Galactic populations,  constitute a wealth of valuable information for theoretical studies. 
In studying the chemical properties of stellar populations in the Galaxy, some of the above works have obtained discrepant results. For instance, assuming Mg instead of Fe as reference element (Mg is thought to come mostly  from core-collapse supernovae, CCSNe), \citet{WEI19} analyzed a spectroscopic sample of 20,485 stars within the SDSS/APOGEE survey. They derived the median trends of abundance ratios [X/Mg] versus [Mg/H] for 15 elements and fit these trends with a simple “two-process” model that characterizes the relative production of CCSNe and SNe\,Ia.
They found, for the particular case of oxygen, that [O/Mg] barely correlates with [Mg/H]. Conversely, \citet{GRI19} found a strong correlation, with [O/Mg] values that significantly decrease with increasing [Mg/H], a result in agreement with the work by \citet{BEN14}. 
 As above mentioned, inferring  oxygen
abundances from  optical spectra  or near-IR spectra  can be very challenging due to, principally NLTE corrections or  the impact of  $T_{\rm eff}$ on OH molecular abundances, respectively.

In this paper,  we take advantage of the GES spectroscopic Survey to derive oxygen abundances for a  sample  of 516 FGK dwarf stars belonging to the Galactic disks. These stars are a sub-sample of a larger (2133 objects) set of dwarfs stars already
analyzed by \citealt{FRA20} (hereafter, FR20) to derive  carbon abundances from atomic lines in their UVES spectra. In this paper, in order to derive oxygen abundance,  
we use the [O\,{\sc i}] 6300.304\,\AA~ forbidden line, which is  unaffected by NLTE effects, with the aim to construct and analyze
the trends of [O/H], [O/Fe], [O/Mg], and [C/O]\footnote{Iron and magnesium abundances are from the fifth Gaia-ESO Survey internal data release and carbon abundances are from FR20.} versus [Fe/H], [Mg/H], and spatial position in the Galaxy and  age for the thin  and thick disk populations. Our final goal is to obtain detailed information on the chemical evolution, in particular, of  [O/Fe] and [C/O].
The sample contains stars with $T_{\rm eff}$   from 4877 to 6561 K and  [Fe/H]  from -0.84 to +0.46\,dex.

The paper is structured as follows:  in Section\,\ref{sec:sample} we introduce the sample used in this work. In Section\,\ref{sec:oxygen} we  describe the method adopted for the oxygen abundance determination and  we define the thin  and thick disk samples by using  the three  selection criteria presented in FR20.
Section\,\ref{sec:OFe} is dedicated to the discussion of the [O/Fe]-[Fe/H] and [O/Mg]-[Mg/H] trends both for the thin and thick disk samples while Section\,\ref{sec:CO} is dedicated to the [C/O] trends with [Fe/H], [Mg/H], and [O/H]. 
We also discuss the  [O/H], [O/Fe], [O/Mg] and [C/O] trends with  $R_{\rm med}$ (the mean of the apo- and pericentric distances of the stellar
orbit), with  $|Z_{\rm max}|$ (the maximum distance from the Galaxy’s plane achieved by a star during its orbit) (Section\,\ref{sec:orbits}), and with age (Section\,\ref{sec:age}).
A comparison of our results  with those from other surveys is presented in Section\,\ref{sec:surveys}.
Conclusions are given in Section\,\ref{sec:conclusions}

\section{Observational data}
\label{sec:sample}
For the aim of our work, we started from the sample of 2133 observed stars extracted from the fifth Gaia-ESO Survey internal data release (GES iDR5) and whose carbon abundances were derived from atomic lines in FR20. The sample (hereafter UVES-U580 sample) contains spectra of FGK dwarf stars obtained with the UVES spectrograph in the setup centred at 5800\,\AA.  These spectra are exposed onto two CCDs, resulting in a wavelength coverage of 4700$\div$6840\,\AA~with a gap of $\sim$50\,\AA~in the center and  a spectral resolving power $\frac{\lambda}{\Delta\lambda}\sim$ 47,000 and $\sim$52,000\footnote{according to the SPEC\_RES keyword in the primary header of the observed spectra.} in the blue and red part of the spectra, respectively.
For the sake of easy reference we here only mention the main properties of our sample. The reader is referred to FR20 for details about the criteria adopted to select the GES spectra.

The UVES-U580 sample consists of  dwarfs stars  with homogeneously  determined GES effective temperature
($T_{\rm eff}$), surface gravity (log\,$g$), iron abundance ([Fe/H]), microturbulence ($\xi$), radial velocities ($v_{\rm r}$), rotational velocities ($v\sin i$) and detailed chemical compositions\footnote{GES abundances of element X are given as:
$\log \epsilon_{X}  = \log \frac{N_X}{N_H} + 12.0$}, spanning the following ranges: 
$T_{\rm eff}$  from 4599 to 6868\,K; log\,$g$ from 3.50 to 4.79\,dex; [Fe/H] from -0.90 to +0.60\,dex; [C/Fe] from -0.44 to +0.44\,dex. In FR20  the sample was restricted to only dwarf and turn off stars in order to avoid the stellar evolution effects in the red giants where during the  first dredge-up phase,  their original CNO atmosphere composition  is altered \citep{IBE64, IBE67, BEK79}. 
 
The availability of accurate parallaxes and proper motions for a large fraction of  stars in our sample\footnote{We considered only stars with small relative errors i.e. less than 10\% in parallaxes and proper motions.}   from the second Gaia data release  \citep[DR2][]{GAI18} and radial velocities from GES iDR5 allowed FR20 to compute Galactic velocities, orbits, and absolute magnitudes for 1804 dwarfs  together with   derived Bayesian ages for 1751 stars. 

\section{Oxygen abundances}
\label{sec:oxygen}
The GES spectra do not cover  the OI 7771-3\,\AA~ triplet lines but incorporate the OI 6158\,\AA~permitted line and  the forbidden  [O\,{\sc i}] 6300 and 6363\,\AA~lines. Since the crowding of spectral features surrounding the stellar OI 6158\,\AA~line makes the required normalization to the continuum a very challenging process, and because of the intrinsic weakness of the feature, after a detailed inspection of the spectra we decided not to use this line. The forbidden [O\,{\sc i}] 6363\,\AA~line  is also very weak in our spectra and not suitable to derive reliable abundances. So, we were compelled to restrict our oxygen abundance determination only to  the strongest forbidden line of oxygen at 6300\,\AA. 

\begin{figure}[htbp]
\includegraphics[width=0.6\textwidth]{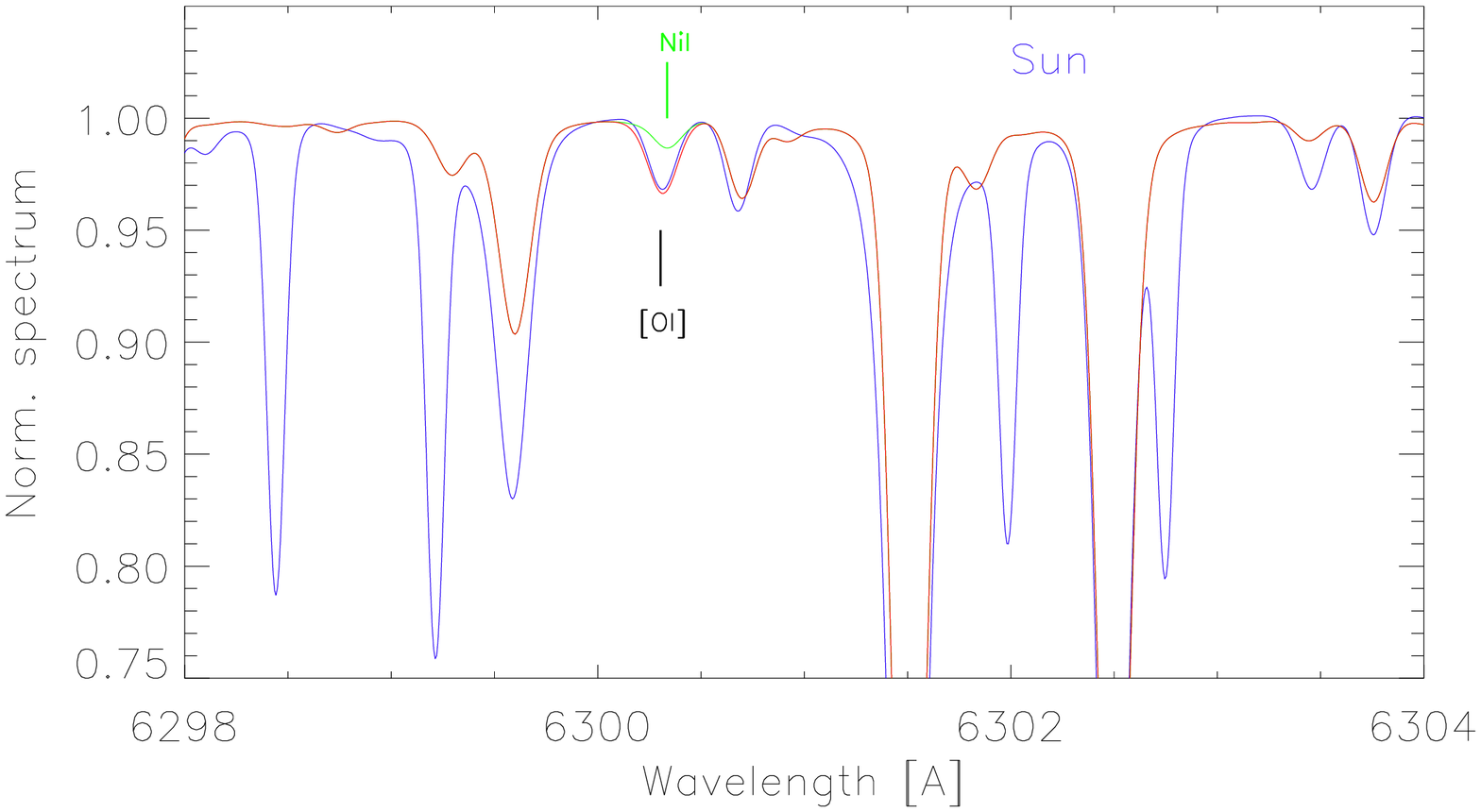}\\
\includegraphics[width=0.6\textwidth]{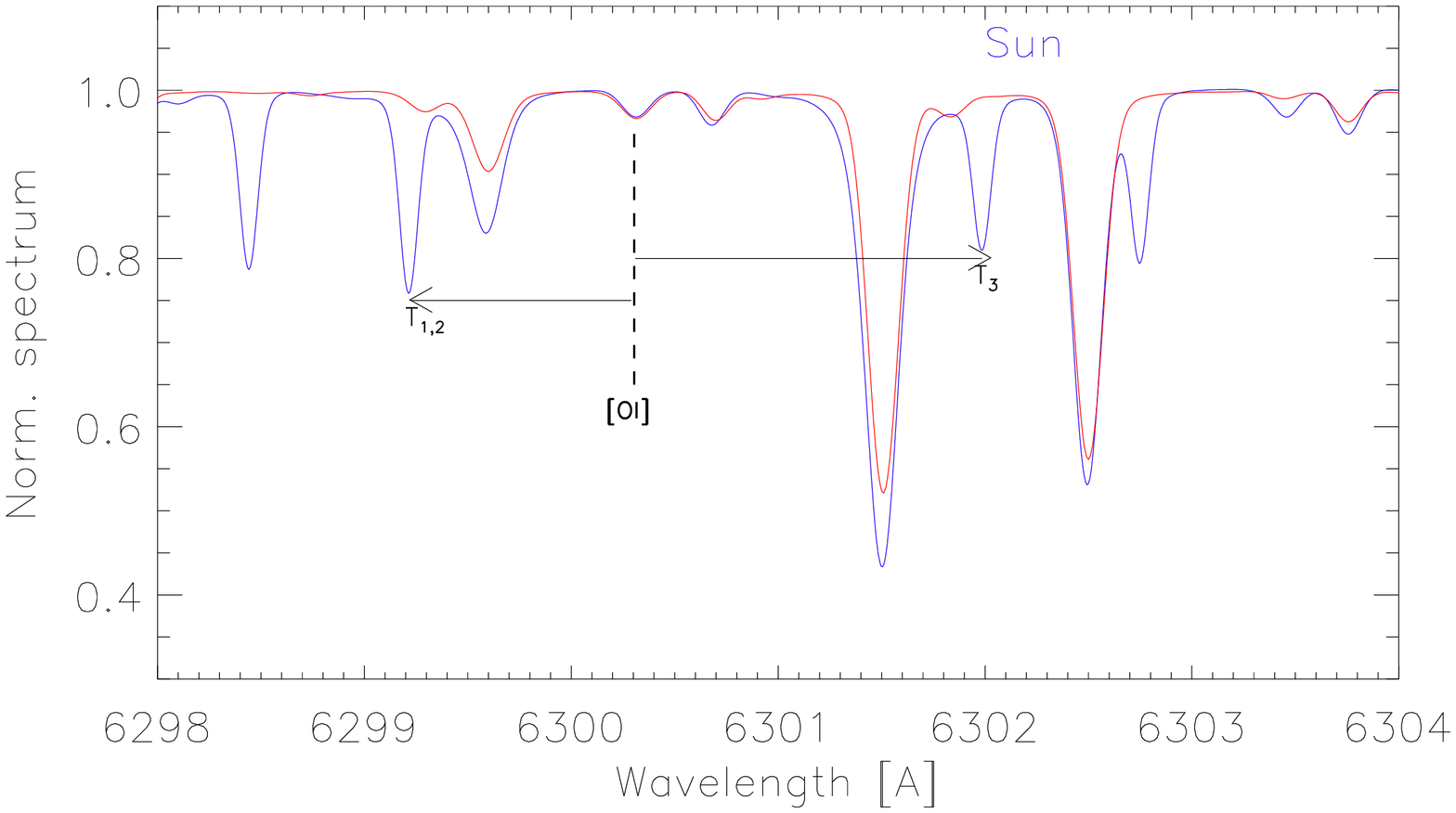}  \\
\includegraphics[width=0.6\textwidth]{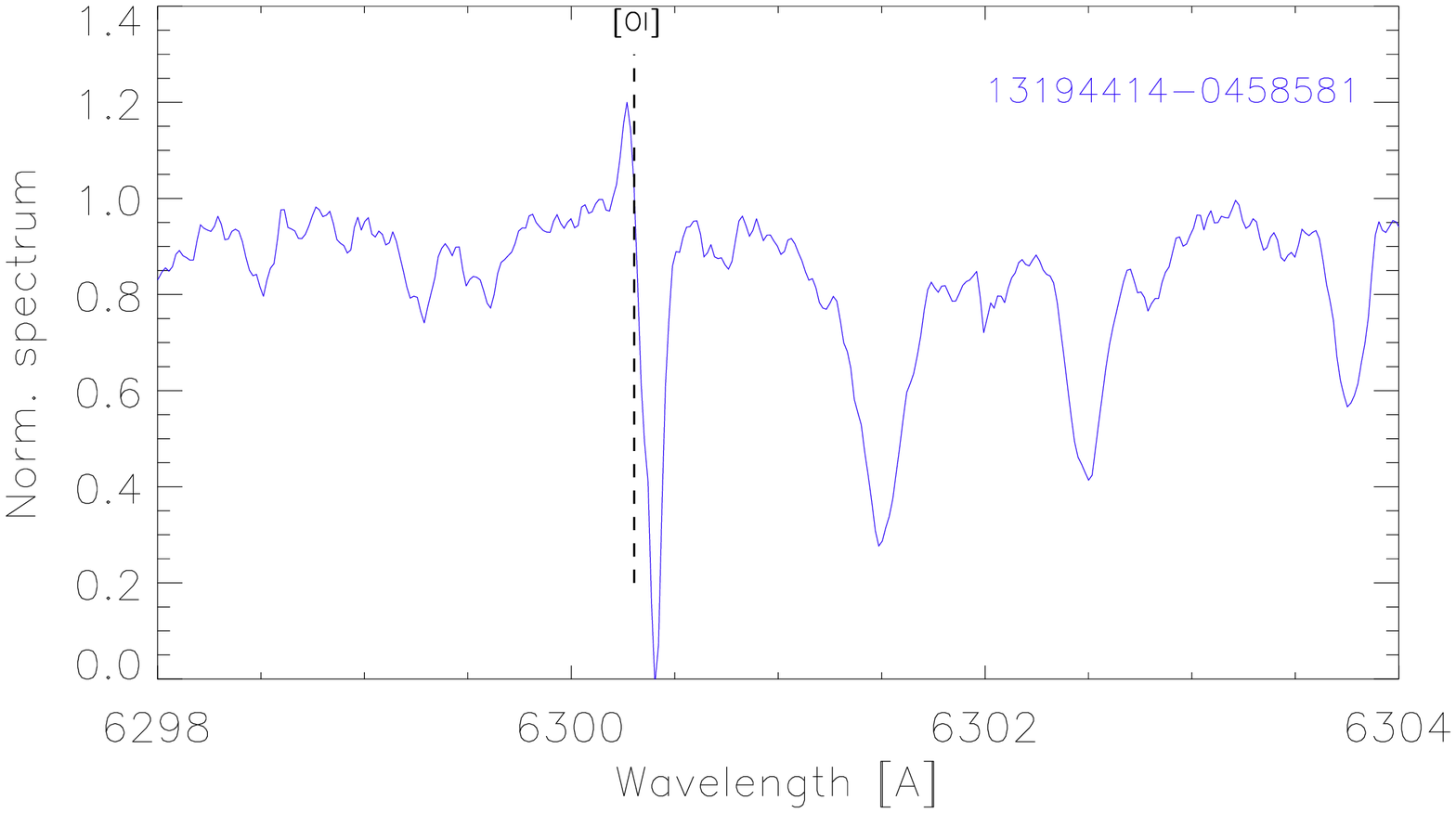}\\
\caption{Top panel: Contribution of  Ni\,{\sc i} lines into the formation of the 6300.3\,\AA~ blend in the solar spectrum:   observed solar spectrum (blue)  and  synthetic spectra  with synthesized both [O\,{\sc i}] and Ni\,{\sc i} lines (red) or without the   [O\,{\sc i}] line (green). Middle panel: comparison of synthetic (red) and observed (blue)  solar spectra; strong observed features not present in the synthetic spectrum correspond to telluric lines: e.g.
T$_{1,2}$ and T$_3$ indicate the 6299.214 O$_2$, 6299.227 H$_2$0, and 6301.985  O$_2$ lines.   Horizontal lines show the shifts of an observed stellar spectrum which would bring the telluric lines T$_{1,2}$ or T$_3$ to overlap the [O\,{\sc i}] feature: the corresponding velocity offsets due to  the combined effect of stellar radial velocity and heliocentric velocity of the observation are -39 and +92\,km/sec, respectively.  Bottom panel: example of GES\,UVES--U580 spectra with residual of the removal of the night-sky [O\,{\sc i}]=6300.30 emission line contaminating the [O\,{\sc i}] stellar feature.
\label{fig:blend}}
\end{figure}

The [O\,{\sc i}] at 6300\,\AA~line is blended with the Ni line at 6300.336\,\AA~(see upper panel of figure\,\ref{fig:blend} and \citealt{BEN04}), and, therefore, the determination of oxygen abundance via this line implies the knowledge of the Ni abundance. Fortunately,  GES iDR5 provides the Ni abundance for all the stars of our sample. 
In order to properly account for the Ni contamination and considering that the line is usually very weak we have used the very high quality  spectrum of the Sun used in FR20 to  fine tune the  log{\it gf} values of all the lines needed to satisfactorily reproduce the [O\,{\sc i}] and Ni\,{\sc i} line blend in the solar spectrum (see top panel of figure\,\ref{fig:blend}).  The adopted
astrophysical log{\it gf}, derived  as described in \citet{FRA18}, are reported in Table\,\ref{tab:loggf} together with values from the literature; we point out that we have  used the same $\log gf$ value (-2.21) for the two isotopic components of the Ni line \citep[see][]{JOH03J} and an isotopic abundance ratio $\frac{{\rm Ni}^{58}}{{\rm Ni}^{60}}=2.6$.

It is important to also note that the [O\,{\sc i}] at 6300\,\AA~line can be, depending on the stellar radial velocity and on the heliocentric velocities of the observations,  blended, sometimes severely,  with telluric absorption lines  as sketched in the middle panel of figure\,\ref{fig:blend} and affected, in some spectra, by poor removal of the night-sky [O\,{\sc i}]=6300.30\,\AA ~emission line  \citep{OST96,BUR77,NIS92} (see the  bottom panel of figure\,\ref{fig:blend}). Therefore, particular care was adopted to check and reject the spectra  where the [O\,{\sc i}] line profile was affected by the above-mentioned problems (see Section\,\ref{sec:OFe_det}).

\begin{deluxetable}{cccl}
\tablecaption{Oscillator strength {\it gf} of [O\,{\sc i}] and Ni\,{\sc i} lines from various sources.\label{tab:loggf}}
\tablewidth{0pt}
\tablehead{
\colhead{Line} & \colhead{[O\,{\sc i}]} & \colhead{Ni\,{\sc i}} & \colhead{References}\\
\colhead {[\AA]} & \colhead{6300.304} & \colhead{6300.336} & \colhead{}}
\startdata
      & -9.717  & -2.310  & 3D, \citet{ALL01}  \\
 & -9.819  & -2.135 & 1D, \citet{BEN04}  \\
$ \log gf$  & -9.717  &  -2.110  & 1D, \citet{BER15} \\
  & -9.717  & -2.602 & 1D, \citet{PAV19} \\
   & -9.715  & -2.210  & 1D,  derived and adopted in this paper \\
\enddata
\end{deluxetable}

\subsection{Atmosphere models and synthetic spectra}
To estimate  [O/Fe] ratio abundances we  adopted a spectral synthesis technique.
For each of the 2133 stars,  we used the stellar atmosphere  ATLAS12  code  \citep{KU05a} 
and the spectral synthesis program SPECTRUM v2.76f \citep{GRA94}  to compute   its model atmosphere and synthetic spectrum, respectively. 
As described in FR20, we adopted 
ATLAS12 since it allows us to generate on-the-fly  models  with full consistency between the chemical composition used to build the atmosphere structure and the one actually used in synthesizing the emergent spectrum. In particular, for each {\it i}-th star, we used its GES iDR5 atmospheric parameter values ($T_{\rm eff}$, log\,$g$,[Fe/H], and  $\xi$) and individual element abundances but for [C/Fe] values, which are from FR20  (for those elements with no estimate of [X/Fe] we assumed [X/Fe]=0) to compute 13 atmosphere models  differing only in [O/Fe], i.e with [O/Fe]$_j=-0.6+(j-1)\times0.1$\,dex (with {\it j}=1,..,13).

Then, starting from each i,j (i and j specify the star and the adopted [O/Fe] ratio, respectively) atmosphere model,   we used
SPECTRUM v2.76f to obtain  the corresponding  normalized spectrum (S$^{\rm i,j}_{\rm N}$) in LTE approximation. 
The line list of atomic and molecular
transitions we used in computing the synthetic spectra is the  INTRIGOSS \citep[high resolution synthetic spectral library,][]{FRA18}\footnote{http://archives.ia2.inaf.it/intrigoss/} line list, updated by the new log{\it gf} values of Table\,1 in FR20,  and extended, by using the same technique described in \citet{FRA18}, to cover also the wavelength range 6280--6325\,\AA. We adopted for the reference
solar abundances those obtained by \citet{GRE07}.

In conclusion, we computed for each star a set of  13  normalized synthetic spectra with  [O/Fe] consistent with the above mentioned atmosphere models. Since 
the synthetic spectra were computed at a very high resolving power ($\sim240,000$), they were broadened by using the GES iDR5  $v\sin i$ stellar values and degraded at the resolution of red UVES spectra  ($R\sim$52,500).

In order to remove the instrumental signature in the [O\,{\sc i}]=6300\,\AA~line region of the observed (stacked) UVES-U580 spectra we used, for each star {\it i}, the {\it j} normalized synthetic S$^{\rm i,j}_{\rm N}$ spectra  
to obtain from the corresponding observed  UVES-U580 spectrum a set of 13 normalized observed  spectra (O$^{\rm i,j}_{\rm N}$). 
The normalization was performed by applying the technique described in \citet{FRA18, FRA20} but  only in  the region surrounding the stellar [O\,{\sc i}] 6300\,\AA~line. 
Actually, we extracted from the observed spectrum the 6280-6325\,\AA~wavelength region, we searched for quasi--continuum flux reference points in S$^{\rm i,j}_{\rm N}$
(i.e. wavelength points with flux levels in excess of 0.97) taking care to exclude regions affected by telluric  features in the observed spectrum, and, then, we used the same points in the corresponding observed UVES spectrum to derive the continuum shape via a linear fitting of the ratio between  observed  and  synthetic spectra. 
Eventually, the observed spectrum is divided by the so computed linear fit to obtain the normalized spectrum O$^{\rm i,j}_{\rm N}$.

\subsection{{\rm [O/Fe]} determination: synthesis of the {\rm [O\,{\sc i}] 6300\,\AA}~line}
\label{sec:OFe_det}
For  each {\it i} star and for each {\it j} pair of spectra, i.e. for different [O/Fe] values, we computed the  standard deviation ($\sigma_j^i$) between O$^{\rm i,j}_{\rm N}$ and S$^{\rm i,j}_{\rm N}$ in a wavelength region centered at 6300.3038\,\AA~ and with a  width  proportional to the stellar rotational velocity ($v\sin i$) taking into account also its uncertainty ($\epsilon_{v\sin i}$)  (i.e. $\lambda_0\pm\Delta\lambda_{\rm rot}$ where  $\lambda_0=6300.3038\,\AA$, ~$\Delta\lambda_{\rm rot}=(v\,sin i+\epsilon_{v\,sin i})\lambda_0/c$, and $c$ is the speed of light). The above process was implemented after we checked  for:
\begin{itemize}
    \item the presence of a badly removed night-sky [O\,{\sc i}]=6300.30\,\AA~ emission line  in the normalized spectra and all the spectra showing  points with normalized flux values greater than  1.05 in the spectral range defined for computing  $\sigma_j^i$ were rejected;
    \item the presence of contaminating telluric lines by rejecting all the spectra with regions 1.5 times below the minimum predicted by all the S$^{\rm i,j}_{\rm N}$ spectra  in the spectral range defined for computing  $\sigma_j^i$.
\end{itemize}
Then, using a parabolic fitting, we determine the ``best'' [O/Fe] value corresponding to the position of the minimum (if any) of $\sigma_j^i$ vs [O/Fe] (see for details FR20, top panels of their Figure\,2). 
In such a way,  we were able  to obtain [O/Fe] estimates  for 869  stars (no clear minimum in $\sigma_j^i$  was detected for the other  stars, thus preventing a sound determination of [O/Fe]). 
We assumed as the uncertainty in the obtained [O/Fe]
the half-step of our [O/Fe] grid of models and synthetic spectra, i.e. $\sigma_{\rm [O/Fe]} \pm 0.05$\,dex.

 To evaluate the variation of the oxygen abundance determinations when the uncertainties in the adopted model atmosphere parameters are taken into account, we re-derived [O/Fe] for some representative stars by varying their input $T_{\rm eff}$, log\,$g$, and [Fe/H] values by plus or minus their typical uncertainties which, for our sample, are 60\,K, 0.1\,dex, and 0.1\,dex, respectively. We choose two pairs of a relatively cool and hot stars one with [0/Fe]$\simeq$0 and one with  [0/Fe]$>$0.3\,dex. In any case, the new [0/Fe] determinations differ from the nominal ones by less than 0.05\,dex and indicated that the most critical parameter is  log\,$g$. Furthermore, to avoid any systematic error due to the use of a possible incorrect   
 Ni abundance in evaluating the blend contribution of the Ni lines, we decided to re-apply our fitting procedure also to only one half of the [O\,{\sc i}] profile  (i.e. the blue wing of the line profile since the blend effect is much stronger in the red part of the [O\,{\sc i}] line profile as shown in the top panel of figure\,\ref{fig:blend}) and to compare the so obtained  [O/Fe] with that derived from the whole profile. Eventually we rejected all stars/spectra  with a standard deviation of the two obtained [O/Fe] values  greater than  $3\sigma_{\rm [O/Fe]}$  thus obtaining a final sample of 516 dwarfs stars with trustworthy [O/Fe]  abundance ratios (the adopted values are those from the fit of the whole profile). 
Moreover, to assess the absence of any other systematic offset in the derived [O/Fe] values we applied  the above-described  procedure also to the solar spectrum obtaining   [O/Fe]$_{\sun}=-0.02$.

The comparison of our results for oxygen abundance determination with those available in the literature can be done, unfortunately, only for very few of our stars. In fact,  we found only 3, 4, and 10 stars in common with \citet{BEN14}, \citet{BRE16}, and  \citet{BUD19}, respectively. In any case, our results are in reasonable agreement with those obtained in these three works taking into account that different oxygen lines and NLTE corrections were used (we recall that our results do not require any NLTE correction since we used the [O\,{\sc i}] 6300 line). Actually,  we found mean differences in [O/H] of 0.06$\pm$0.08, -0.24$\pm$0.10, and 0.09$\pm$0.12 for the stars in common with  \citet{BEN14}, \citet{BRE16}, and  \citet{BUD19}, respectively, where the highest difference  is very likely due to the absence of any NLTE correction in \citet{BRE16}.

Out of the 516 stars 308  have also an age determination (for the other 208 stars FR20 found too uncertain ages, i.e. with FWHM$_{age} > 8$\,Gyr).    

\begin{figure}[ht!]
\plotone{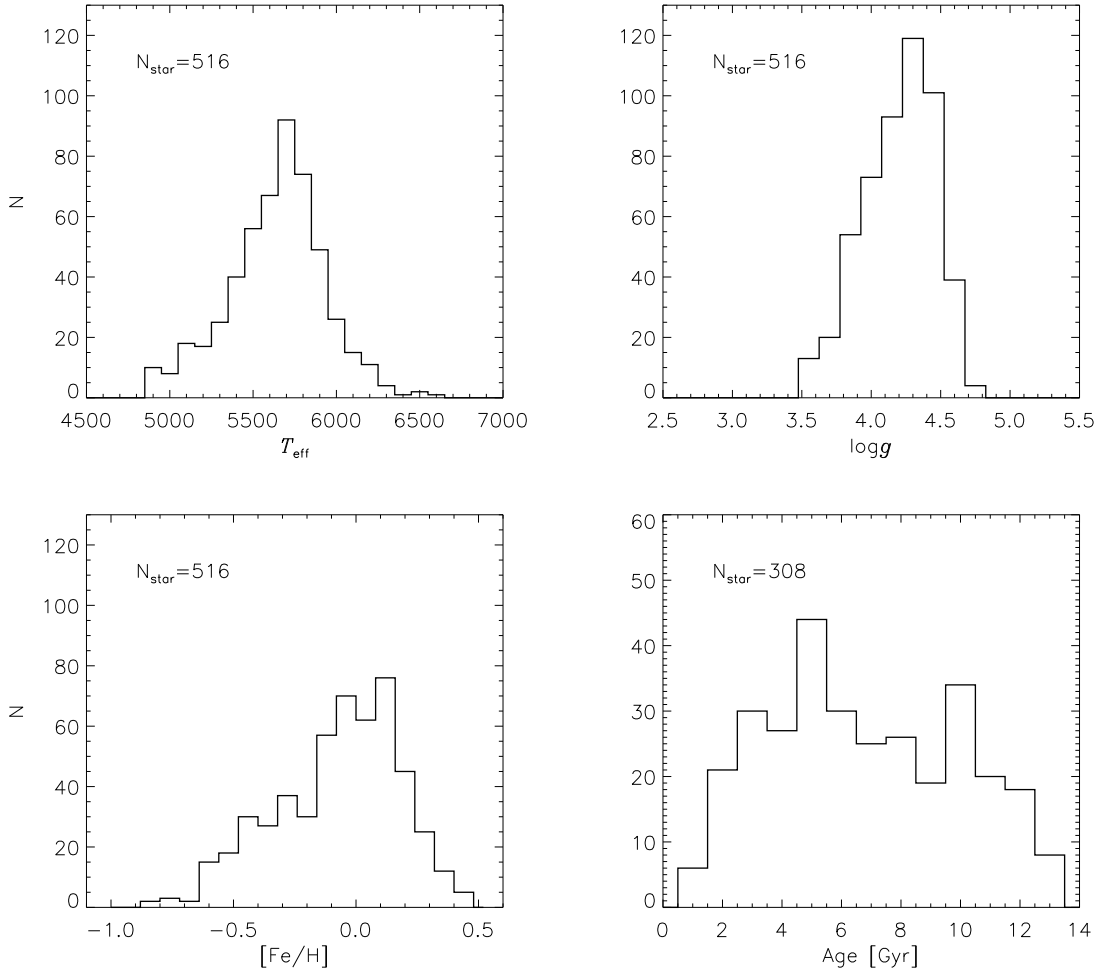}
\caption{ Histograms
of atmospheric parameter values for the  516  stars with trustworthy [O/Fe]: $T_{\rm eff}$ (top-left panel) log\,$g$ (top-right panel),    [Fe/H] (bottom-left panel), and of age (bottom-right panel) for the sub-set of 308 stars (see text).  \label{fig:param}}
\end{figure}

Figure\,\ref{fig:param} shows the  histograms  of the atmospheric  parameters and FR20 ages for the stars in the final sample.  
The final sample consists of   stars spanning in $T_{\rm eff}$  from 4877 to 6561\,K; log\,$g$ from 3.51 to 4.79\,dex; [Fe/H] from -0.84 to 0.46\,dex.

\subsection{Definition of thin and thick disk star samples}
\label{samples}

To discriminate between thin and thick disk stars in the UVES--U580 sample we used, as in FR20, three different selection methodologies: {\it (i)} the chemical one based on positions in the [Mg/Fe]-[Fe/H] plane; {\it (ii)} the kinematical one  based on stellar Galactic velocities; {\it (iii)} the orbital one based on the  $R_{\rm med}$  and  $|Z_{\rm max}|$ orbital parameters (see FR20  for details).  As already discussed in FR20 (and references therein) there is not yet conclusive criteria for unambiguously separating samples of thin and thick disk stars and with null contamination. In fact, depending on the adopted criterion, the number of classified (i.e. as thin or thick disk stars) and of unclassified stars in  our sample actually varies. 
 In fact, even if thin and thick disk stars should differ both in chemistry and kinematics, the sensitivity of each above listed classification method varies for different kind of stars. In particular:
\begin{itemize}
\item the adopted chemical classification, taking also into account the uncertainties on [Mg/Fe] and [Fe/H] determinations, is less effective for metal rich stars due to the merging of the two sequences of high- and low-$\alpha$
objects at [Fe/H]$>$0;
\item the kinematical classification, also affected by uncertainties in the computed Galactic velocities, has difficulties in classifying objects which fall in the overlap regions of the Gaussian velocity distributions of the different Galactic components;
\item the classification based on the orbital parameters $R_{\rm med}$  and  $|Z_{\rm max}|$, whose computed values depend not only by the accuracy of  the input stellar positions, distances, and velocities, but also by the reliability of the adopted Galactic potential, may fail for stars falling close to the somewhat arbitrary separation borders in the orbital parameter plane between the different Galactic components adopted in FR20.
\end{itemize}

On the other hand,  FR20 showed that, independently of the classification method adopted, in all the  thin and in all the thick disk star sub-samples the behaviours  of [C/H], [C/Mg], and [C/Fe]  versus [Fe/H], [Mg/H], and age and the Galactic regions they populate (given by the stellar orbit $R_{\rm med}$  and  $|Z_{\rm max}|$) are very similar. Therefore, in this paper, we decided to merge the results of the three above-mentioned selection methodologies  in the following way:

\begin{center}
\begin{tabular}{ll}
Thin-disk star if  &  (Thin$^{\rm C}$=true and/or  Thin$^{\rm K}$=true and/or Thin$^{\rm O}$=true) and  \\
& (Thick$^{\rm C}$=false and  Thick$^{\rm K}$=false and Thick$^{\rm O}$=false) \\
Thick-disk star if  &  (Thick$^{\rm C}$=true and/or  Thick$^{\rm K}$=true and/or Thick$^{\rm O}$=true) and  \\
& (Thin$^{\rm C}$=false and  Thin$^{\rm K}$=false and Thin$^{\rm O}$=false) \\
\end{tabular}
\end{center}
where the superscripts ``C'', ``K'', and  ``O'' refer to  the results of the chemical, kinematical, and orbital methodology, respectively.
In such a way we obtained, after discarding the few stars with discordant classification, two samples of 376 and of 20 stars classified by at least one methodology as thin or thick disk stars, respectively even if unclassified by the other methodologies.
In such a way we tried also to deal with the uncertainties on the different parameters used for the different classification approaches. We recognize that the thick disk sample is small and, therefore, it is desirable to collect more data of this population to confirm the robustness of the results presented in the following sections.


 \begin{figure}[htbp]
\includegraphics[width=1\textwidth]{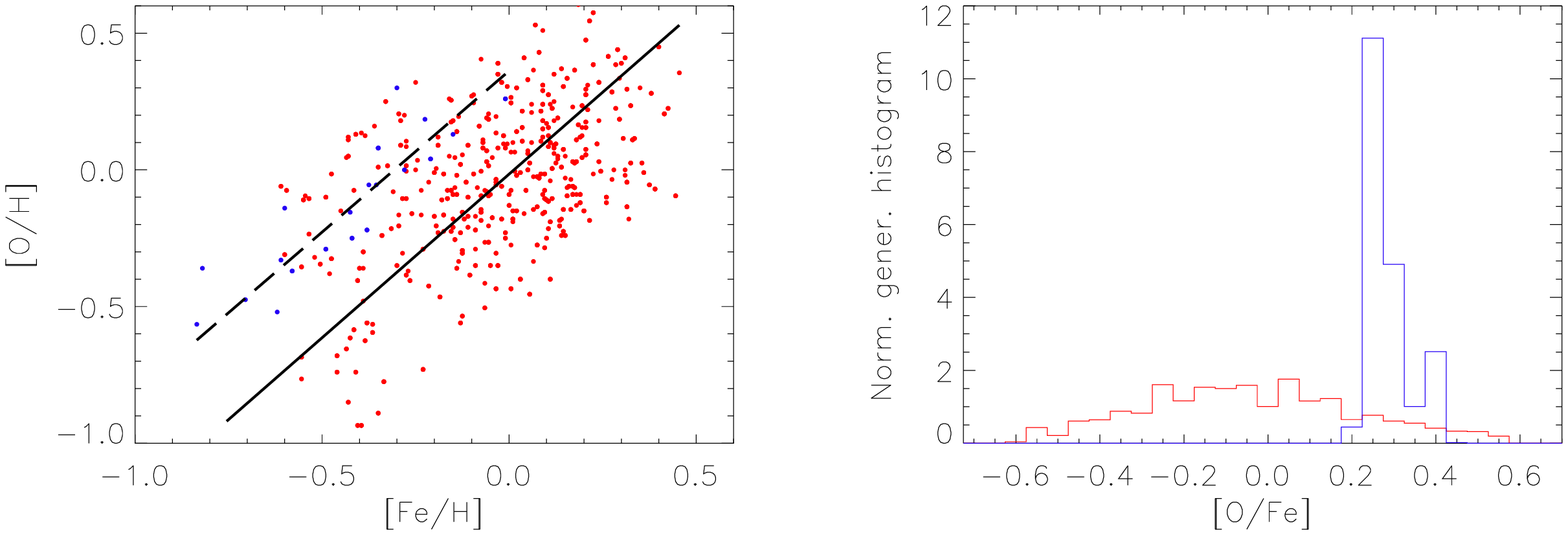}
\caption{[O/H]--[Fe/H] diagrams  (left panel) and  [O/Fe] normalized generalized histograms (right panel) for  thin (red) and thick (blue) disk star samples.
Regression lines for thin (continuous)  and thick (dashed) disk stars are superimposed on the left panel.  Oxygen abundances are from this paper while [Fe/H] are from GES iDR5.
\label{fig:OH}}
\end{figure} 

\begin{deluxetable}{lcrrc}
\tablecaption{Linear regression coefficients  \label{tab:regression}}
 \tablecolumns{6}
 \tablewidth{0pt}
 \tablehead{
 \colhead{} &
\colhead{$s \pm \sigma_s$} &
\colhead{$i \pm \sigma_i$} &
\colhead{~~N$_{\rm star}$} &
\colhead{  $\chi_{\rm r}$}
}
 \startdata
 ~~~~~Thin~~~~~ & 1.20$\pm$ 0.03 & -0.02$\pm$ 0.01  & 376 & 2.9 \\
 ~~~~~Thick~~~~~& 1.18$\pm$ 0.11 & 0.36$\pm$ 0.05  & 20 &  3.5  \\
 \enddata
 \tablecomments{N$_{\rm star}$ is the number of points used to derive the regression lines.}
 \end{deluxetable}

\section{Oxygen trends in the thin and thick disks}
\label{sec:OFe_result}

The left panel of Figure\,\ref{fig:OH} shows the [O/H] vs [Fe/H] for the thin (red points) and thick (blue points) disk sample stars together with their regression lines (continuous and dashed lines for thin and thick disk samples, respectively). 
Even if the thick disk sample does not contain a very large number of stars, we can observe that, in general, the thick disk stars have larger O abundance, by about 0.35\,dex, than the thin disk stars at the same [Fe/H]. The slopes of the regression lines do not differ significantly as can be seen in Table\,\ref{tab:regression} where  the slope ($s$) and intercept ($i$) coefficients with their standard deviations ($\sigma_s$ and $\sigma_i$, respectively) are reported.
 In the right panel of Figure\,\ref{fig:OH} we show the normalized generalized distributions of [O/Fe] built  by summing individual unit area Gaussian computed for each star by using its [O/Fe] value and its uncertainty and normalizing the results to the number of entries. In computing the regression lines and in building the normalized distributions we weighted each star 1, 2, or 3 times according to the  number of methodologies which lead to its classification. 
 The thin disk sample (red) distribution of [O/Fe] is quite broad and has its maximum at  [O/Fe]$\approx$0.0\,dex while the thick disk sample distribution (blue) is much narrower and peaks at  [O/Fe]$\approx$0.25\,dex, being consistent with a much faster and efficient formation of the thick disk.

Figure\,\ref{fig:OH} also  shows that both the thin and thick disk stars display a large scatter in the [O/H] values. 
Such a scatter for this abundance ratio was also found (and discussed) by several authors \citep[e.g.][]{BER15, BER16, 
NIS18, AMA19}, as well as for other chemical elements in Galactic abundance trends, however, to a lesser extent
\citep{ADI12}. Since the above works and ours are based on different oxygen features, the results cannot be easily compared. The quite large vertical extension at a given [Fe/H] might represent a truly cosmic scatter \citep[e.g.][]{PET11, BER16} or its combination with uncertainties in our [O/Fe] determinations.  We, nevertheless, want to remark that if the scatter of the np points was purely statistical, we would expect our fits to have a reduced  $\chi_{\rm r}$=$\sqrt{\frac{\chi^2}{{\rm np}-2}} \lesssim 1$, but our fits have
reduced $\chi_{\rm r} \sim 3$ (see Table\,\ref{tab:regression}), which
suggest that a significant part of the observed scatter 
is astrophysical. 


\begin{figure}[htbp]
\includegraphics[width=0.5\textwidth]{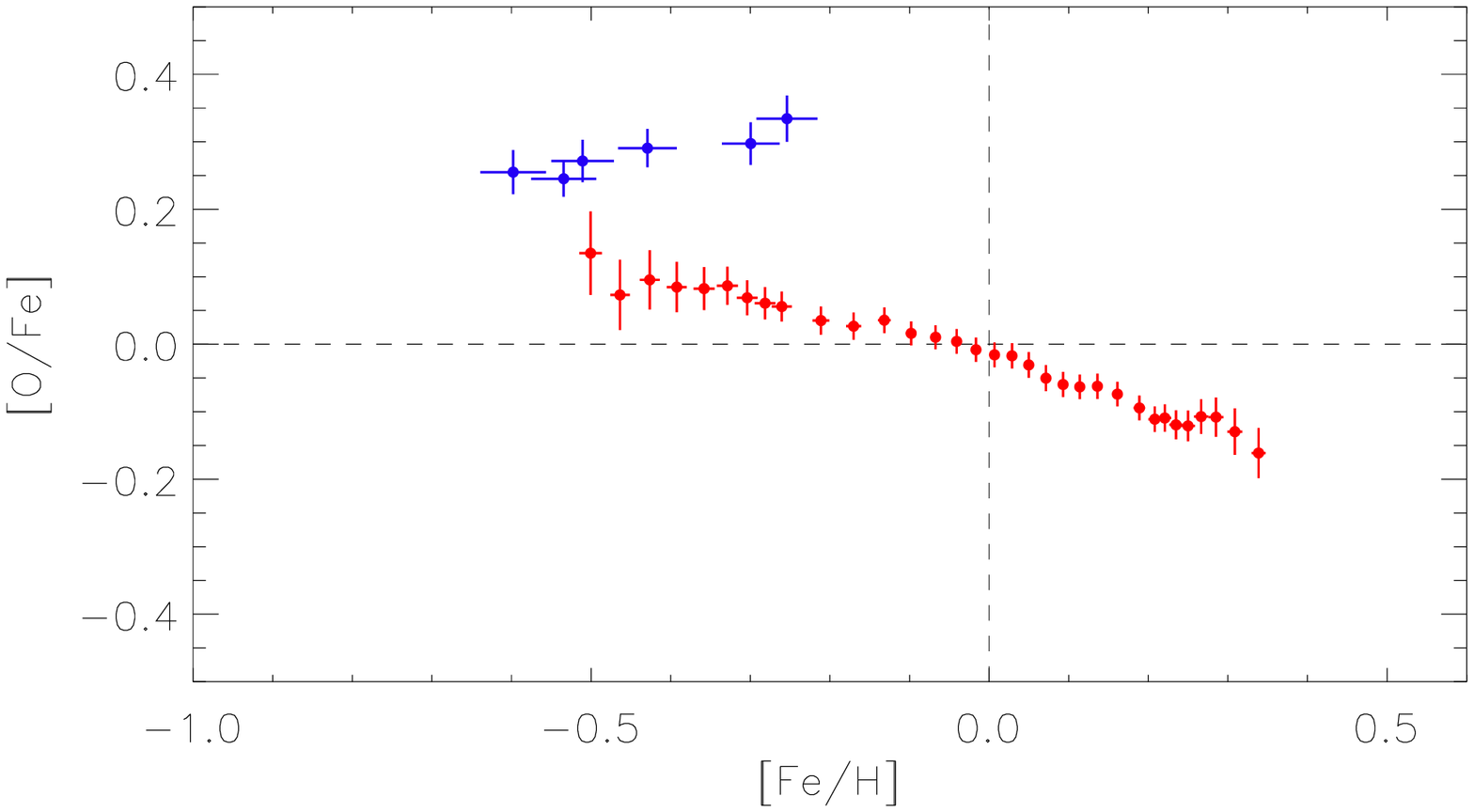}
\includegraphics[width=0.5\textwidth]{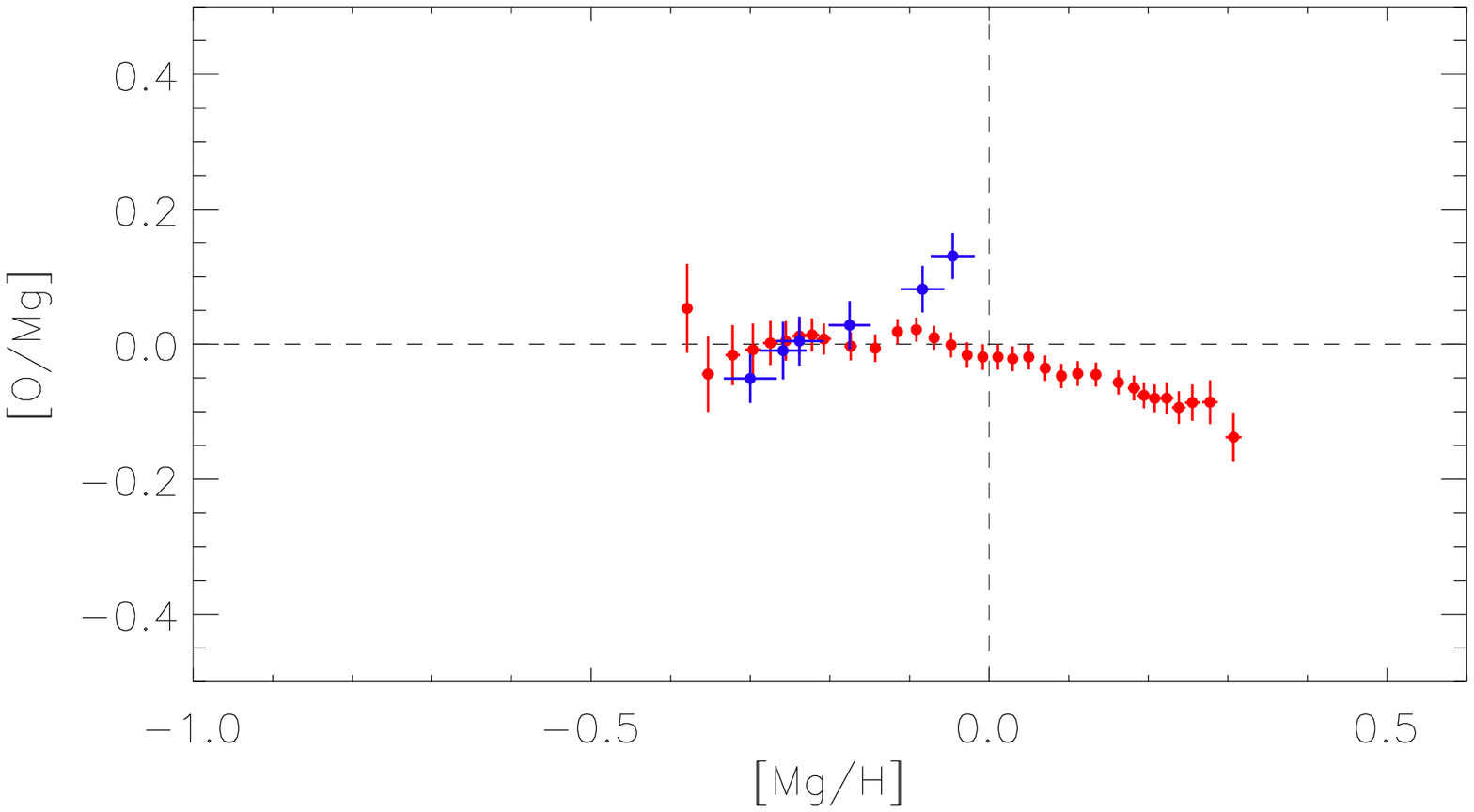}
\caption{[O/Fe]-[Fe/H] diagrams  (left panel) and   [O/Mg]-[Mg/H] diagrams  (right panel) for  thin (red) and thick (blue) disk star samples: binned running averages and standard deviations are plotted.  Oxygen abundances are from this paper while [Fe/H] and [Mg/H] are from GES iDR5.
\label{fig:OFe_Mg}}
\end{figure}

\subsection{ {\rm [O/Fe]-[Fe/H]} and {\rm[O/Mg]-[Mg/H]} trends}
\label{sec:OFe}
As mentioned earlier, several  studies on the Galactic chemical history make use of Mg instead of the more popular and accessible Fe as a reference element. The choice is based on the potential single origin (SNII) of Mg \citep[e.g][]{BEN04,WEI19,GRI19}, although some stellar yield models have provided theoretical evidence that Mg might also be partially released into the interstellar medium by SNIa \citep{MAG17, NAI18}. In this context \citet{BEN04} recommended to be very cautious about promoting one specific reference element and they rather suggested to consider more than one. Following their suggestion, on what follows we analyze disk stellar populations on the basis of the abundance ratios [O/Fe] and  [O/Mg] in terms, respectively, of [Fe/H]  and  [Mg/H]. The trends depicted in Figure\,\ref{fig:OFe_Mg} are computed using a  running average (using bins with 100 points partially overlapped by shifting them by 10 points and bins with 10 points partially overlapped by shifting them by 2 points for the thin and thick disk stars, respectively) while the {horizontal and} vertical bars correspond to the standard deviations of the bin averages.

For the [O/Fe]-[Fe/H] tendencies illustrated in the left panel of  Figure\,\ref{fig:OFe_Mg} we distinguish the following properties:
\begin{itemize}
    \item we found systematic differences between thin  and thick disk stars as already hinted by the distributions in  Figure\,\ref{fig:OH}:  thick disk stars show a higher [O/Fe] than objects in the thin disk in the common [Fe/H] interval. Thick disk stars show an almost flat distribution in the limited [Fe/H] range they cover, while thin disk stars show a clear monotonic decrease of [O/Fe] for increasing metallicity.  These  [O/Fe] vs [Fe/H] trends resemble, as expected, the behavior of other $\alpha$--elements versus [Fe/H]. Actually, being oxygen an $\alpha$--element, an [O/Fe] plateau reflects, for the most metal poor stars in our sample, a steady rate of oxygen and iron enrichment of the ISM  by core-collapse supernovae. The gradual decrease of [O/Fe] with increasing [Fe/H] reflects the iron enrichment from Type-Ia supernovae, which have a delayed distribution in time relative to core-collapse supernovae \citep[e.g.][]{CAR05,MAO12}.  
    Analogous  systematic  differences between  thick and thin disk abundance distributions have been obtained in several works in the literature 
  \citep[e.g.][]{BEN04,RED06,RAM13,BEN14,NIS14,BER15,DEL19}, all confirming that  oxygen  behaves like Mg and the other $\alpha$-elements, but at high metallicity;  
    \item  we can notice that at [Fe/H]$\approx$ -0.40  there is a difference of $\Delta$[O/Fe]$\sim$+0.2\,dex between the thin and thick disk star trends. Interestingly, \citet{RED06} found, on the basis of a  kinematical selection, that their thick and thin disk samples are also well separated. In their bin centered at [Fe/H] = -0.50 the
    mean abundance ratios [O/Fe] are $0.36\pm 0.19$\,dex and 0.24$\pm 0.07$, for the thick  and thin disks, respectively.
    These values are, within uncertainties, in agreement with our results.
    
       In this context, the more recent work of \citet{NIS14}, found  that their thin disk stars fall well below the thick disk counterparts in the [O/Fe] vs [Fe/H] diagram by as much as [Fe/H]$\sim -0.3$ and they  suggested that the two populations merge at higher metallicities. This later feature cannot be corroborated with our data since we lack  thick disk stars with metallicities higher than -0.2\,dex; 
       
    
    \item  at super-solar metallicity ([Fe/H]$>$0) we observe that  the [O/Fe] of thin disk stars  continues to decrease as  also found by several authors \citep[e.g.][]{CAS97,CHE03,BEN04,ECU06,BEN14,AMA19}  in concordance with  the prediction of Galactic chemical evolution models \citep[e.g.][]{CHI03}. In contrast, 
    other works have found a flattening of the trend, i.e. a constant value of [O/Fe]$\sim$0
    \citep[e.g.][]{NIS92,NIS02,RAM13,BER15}. Our results suggest that the oxygen is produced only by CCSNe with no  evidence of  any SN\,Ia or AGB star contribution that would produce a flattening of [O/Fe] at [Fe/H] $\simeq$ 0 as  observed in other $\alpha$-elements  \citep{BEN04,BEN14}.
    We notice that the thin disk star [O/Fe] trend continues  down to an under-abundance of [O/Fe] $\approx$ -0.15 at [Fe/H]$\approx$0.30  in agreement with the results from  \citet{BEN04} who,  adopting kinematic selection for their thin disk sample, show a decreasing trend leading  to [O/Fe] $\approx$ -0.2 at [Fe/H] $\approx$ 0.4\,dex;
    
    \item the linear relations between [O/Fe] and [Fe/H] for the thin disk stars: 

$$\rm {[O/Fe]}  =  (-0.33 \pm 0.01) \times \rm{[Fe/H]} - (0.027 \pm 0.002) $$ 

is in good agreement with the  linear regression obtained by \citet{GUS99} using the abundances of F and G dwarfs by \citet{EDV93}, who found

$$\rm{[O/Fe]} = (-0.36 \pm 0.02) \times \rm{[Fe/H]} - (0.044 \pm 0.010), $$

and plausibly reflects that iron enrichment by Type\,Ia supernovae takes place in larger time scales when compared  with the rapid production of oxygen by  Type\,II supernovae \citep[as first suggested by][]{MAT86};
    \item the thin disk sequence at [Fe/H]=0 is slightly below  zero.  
       We found at [Fe/H]=0 a value of [O/Fe]=$-0.027 \pm 0.002$ which is, in absolute value, slightly larger than the value we derived for the Sun (see Section\,\ref{sec:OFe_det}). Similar results were found by \citet{MEL09} (an average $<$[O/Fe]$>$ = -0.033 $\pm$ 0.011 for 11 stars), and \citet{RAM09} ($<$[O/Fe]$>$ = -0.015 $\pm$ 0.006 for 22 stars) who used samples of 
    solar twins. Similar studies based on F, G and K type stars in the solar neighborhood have also been conducted with compatible results, e.g. \citet{GUS99} ($<$[O/Fe]$>$ =-0.044 $\pm$ 0.010 for 57 stars), \citet{RAM13} ($<$[O/Fe]$>$ =-0.02 $\pm$ 0.04 for 47 stars), and  \citet{NIS14} ($<$[O/Fe]$>$ =-0.016 $\pm$ 0.04 for 11 stars). A possible explanation for the observed
    offset can be found in \citet{NIE12} who suggest that the Sun was born
    at a distance around 5-6\,kpc from the Galactic centre, where  higher metallicity values were reached earlier in the cosmic history of the Galaxy, and it has migrated outwards during its lifetime of 4.5\,Gyr at the current position. An alternative explanation for this peculiarity
    is related to the formation of planetary systems like the solar one and, in particular, to the formation of terrestrial planets \citep{MEL09}.
 \end{itemize}

As far as the [O/Mg]-[Mg/H] trend is concerned (right panel of  Figure\,\ref{fig:OFe_Mg}), we  can make the following considerations:

\begin{itemize}
    \item the decrease of [O/Mg] with [Mg/H]$>0.0$ for the thin disk stars gives evidence  of a different behaviour of oxygen and magnesium  at high metallicities, i.e. they do not evolve in lockstep,  and indicates that, while oxygen is only enriched by CCSNe, magnesium  might be released, to some extent, also by SNIa and/or AGB stars \citep{MAG17, VEN18};
    
  \item  both the thick  and the thin disk stars,  at [Mg/H]$\lesssim -0.1$  show trends almost flat with [Mg/H] at [O/Mg]$\simeq 0$.  These flat [O/Mg] trends are in agreement with the results by \citet{BEN04} (see their Fig.\,12) even if their thick disk values at [Mg/H]$\lesssim $0.1 are almost leveled out   at a  level slightly higher than their thin disk data. The increase in [O/Mg] for the thick disk stars with [Mg/H]$\gtrsim -0.1$ may be due to the poor statistic,  to a selection effect, and to the presence of an  [O/Mg] gradient with the star distance from the Galactic plane as also suggested by the results presented below.
   In fact, since the two bins with higher [Mg/H] contain mainly stars at low $|Z_{\rm max}|$, their high [O/Mg] values may be  simply due to the the negative vertical gradient of this abundance ratio with  $|Z_{\rm max}|$ shown in the third-right panel of Figure\,\ref{fig:OH_Rmed_Zmax}.  
  Therefore, at present it is premature to draw any conclusion. 
 \end{itemize} 
 

\begin{figure}[htbp]
\begin{center}
\begin{tabular}{ccc}
\hspace*{-1cm}\includegraphics[width=0.35\textwidth]{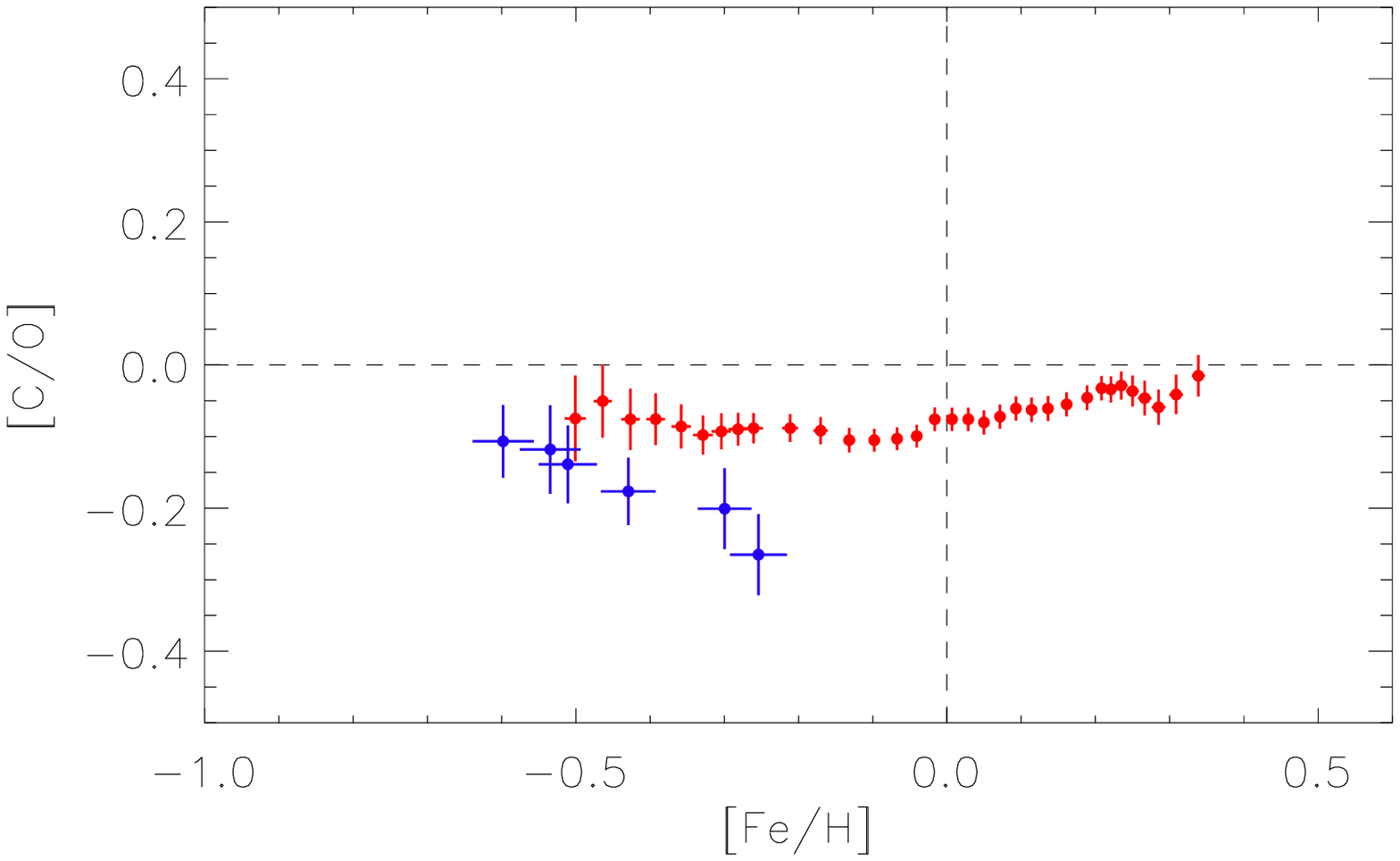}  &
\hspace*{-0.5cm}\includegraphics[width=0.35\textwidth]{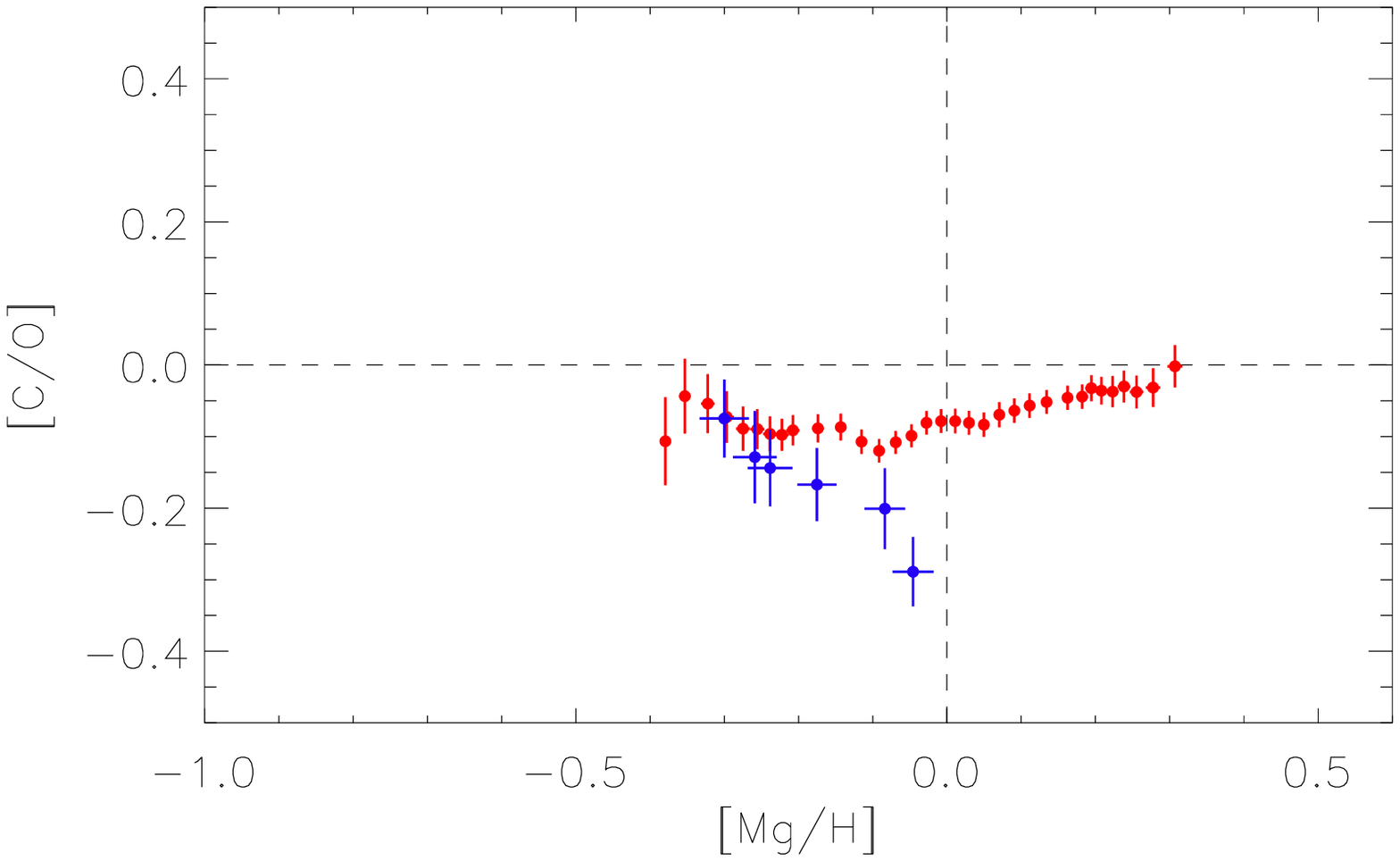}  &
\hspace*{-0.5cm}\includegraphics[width=0.35\textwidth]{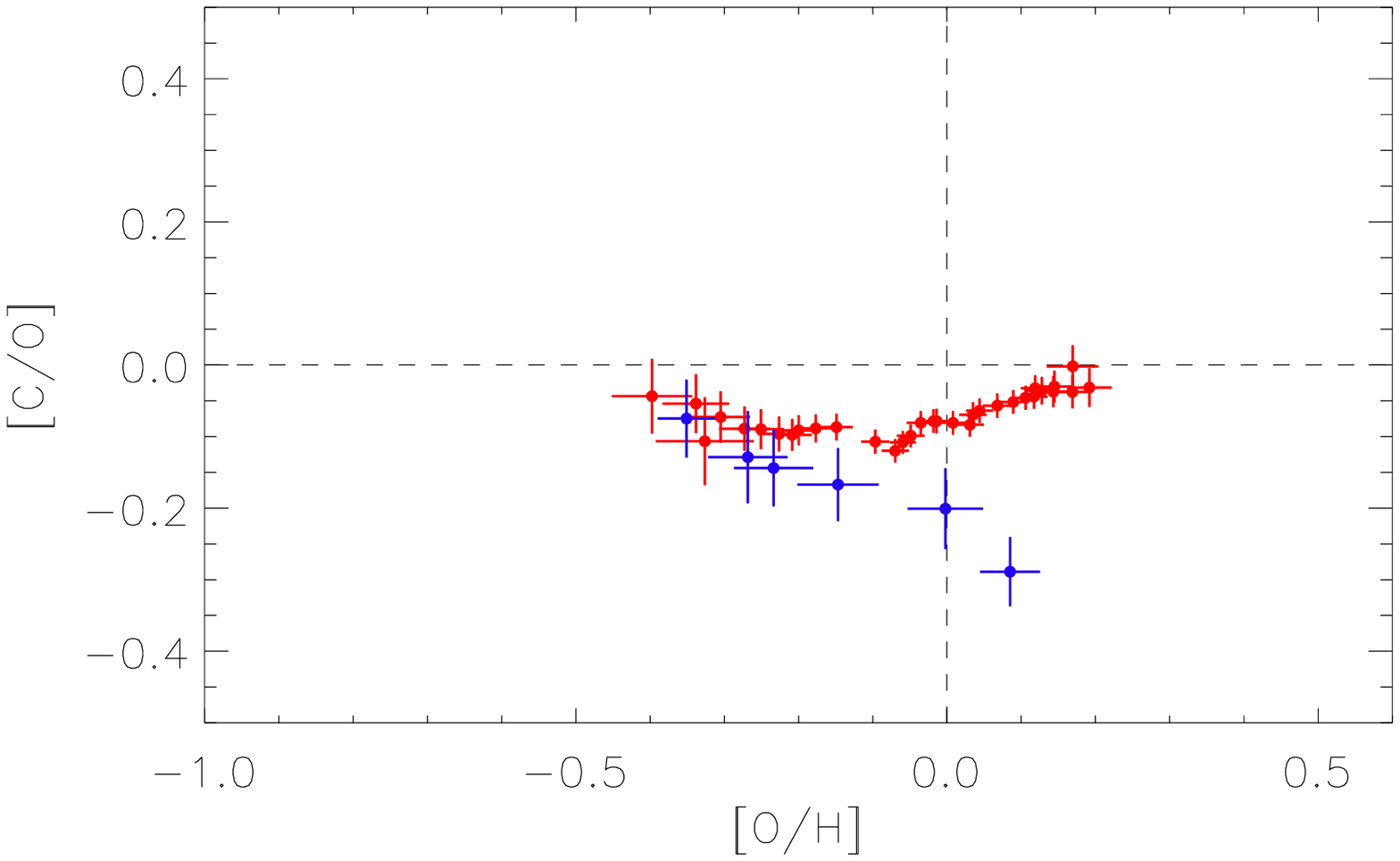}  \\[-10pt]
\end{tabular}
\caption{[C/O] versus [Fe/H] (left panel),  versus [Mg/H] (central panel) , and 
versus [O/H] (right panel) diagrams for  the thin (red) and thick (blue) disk samples.  Oxygen abundances are from this paper, 
carbon abundances are from FR20, and [Fe/H] and [Mg/H] are from GES iDR5.
\label{fig:CO}}
\end{center}
\end{figure} 

\subsection{The {\rm [C/O]} trends with {\rm [Fe/H]}, {\rm [Mg/H]}, and  {\rm [O/H]}}
\label{sec:CO}
The [C/O] ratio in the Galactic disks has also been under debate and is highly relevant in a number of astrophysical scenarios, for example, to understand planet formation processes and the possible existence of terrestrial planets \citep[e.g.][]{DEL10,NIS13, NIS14,BRE17,AMA19}. In the context of this paper, it is particularly useful when discussing the origin and evolution of carbon in the Galaxy. Since  oxygen is exclusively produced in massive stars on a relatively short timescale ($\sim$10$^7$ yr), the change in [C/O] as a function of [O/H] depends on the yields and timescales of carbon synthesized  in different types  of stars 
\citep{CHI03,AKE04,CAR05,CES09,ROM20}.


Figure\,\ref{fig:CO} shows how [C/O] changes
as a function of [Fe/H], [Mg/H], and [O/H] 
for the thin  and thick disk stars. In particular from Figure\,\ref{fig:CO} we can observe:
\begin{itemize}
    \item a slight increment of [C/O] with [Fe/H], [Mg/H], or [O/H] for thin disk stars for ${\rm [Fe/H]}\gtrsim -0.1$.   Such an  increase can be explained taking into account a C production contribution from low–mass stars and/or  from massive stars at high metallicities due to their enhanced mass loss;
\item thick disk stars show  a decrease of [C/O] with [Fe/H], [Mg/H], and [O/H]. A qualitatively similar behaviour can be observed in the [C/O] versus [O/H] correlation of \citet[][see their Figure\,11]{STO20}, even if they interpreted it as due to the presence of a shift ($\sim -0.3$\,dex) of their thick disk star sequence with respect to that for the thin disk. On the other hand, \citet{BEN06} found a flat behaviour for the thick disk stars until [O/H]$\simeq$0
and then an increase. In any case, we should point out  that the thick disk star bins behaviour in our case could be an artifact. Therefore, it would be desirable to expand the thick disk sample, aiming at verifying such a [C/O]  decrease and to elucidate on the effects of the steep vertical gradient of [O/Fe] (see Figure\,\ref{fig:OH_Rmed_Zmax}) on the observed tendency. 
\end{itemize}

\subsection{{\rm [O/H], [O/Fe], [O/Mg]} and {\rm [C/O]} trends with $R_{\rm med}$ and $|Z_{\rm max}|$ }
\label{sec:orbits}

In this Section, we present the trends of oxygen abundance ratios with the mean Galactocentric distance, $R_{\rm med}= 0.5\times(R_{\rm min}+R_{\rm max})$, and with the
maximum absolute distance from Galactic plane,  $|Z_{\rm max}|$ for the 333 thin disk and the 18 thick disk stars for which stellar orbits were calculated in FR20.

\begin{figure}[htbp]
\begin{center}
\begin{tabular}{cc}
\hspace*{-0.5cm}\includegraphics[width=0.45\textwidth]{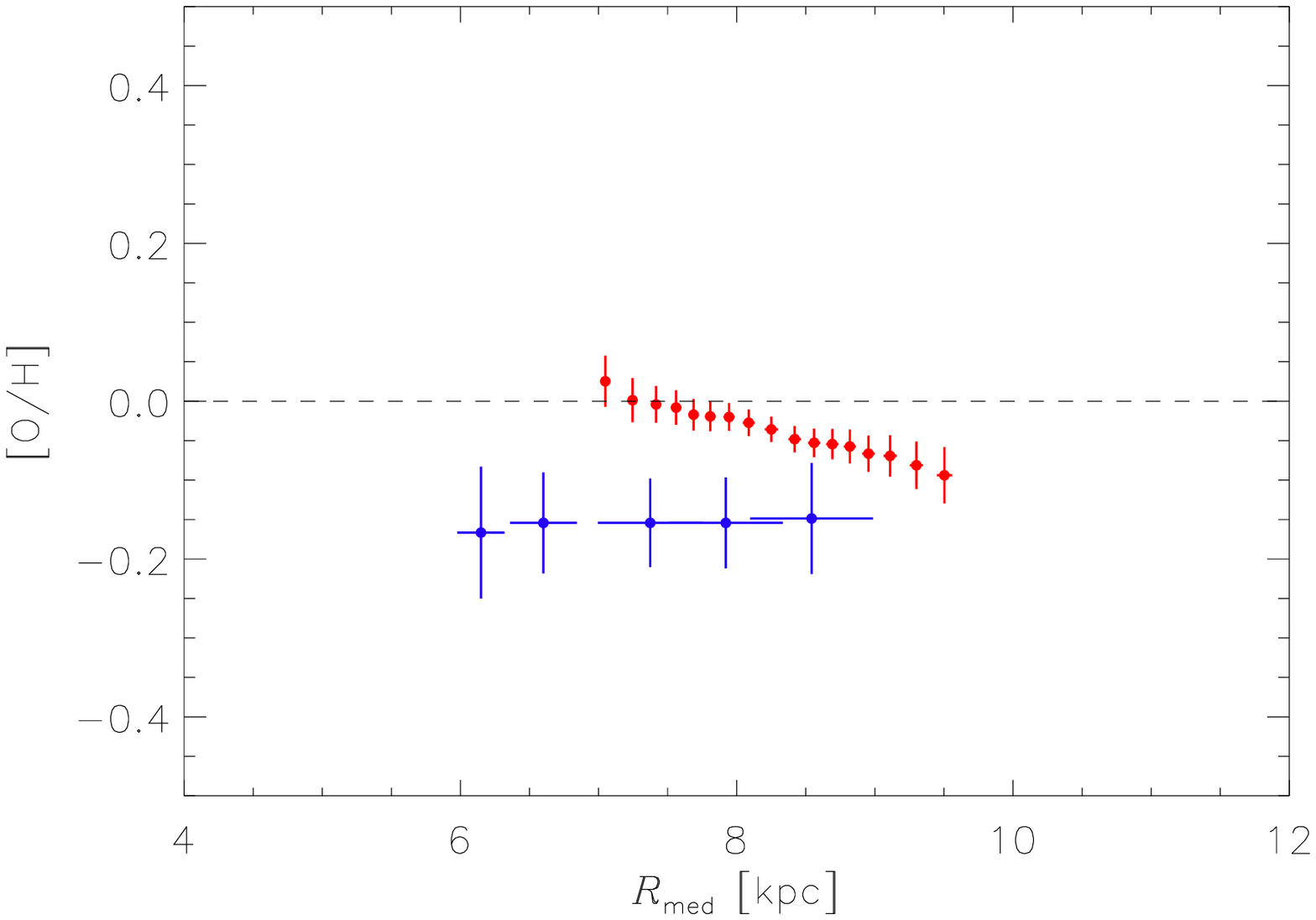} &
\hspace*{-0.5cm}\includegraphics[width=0.45\textwidth]{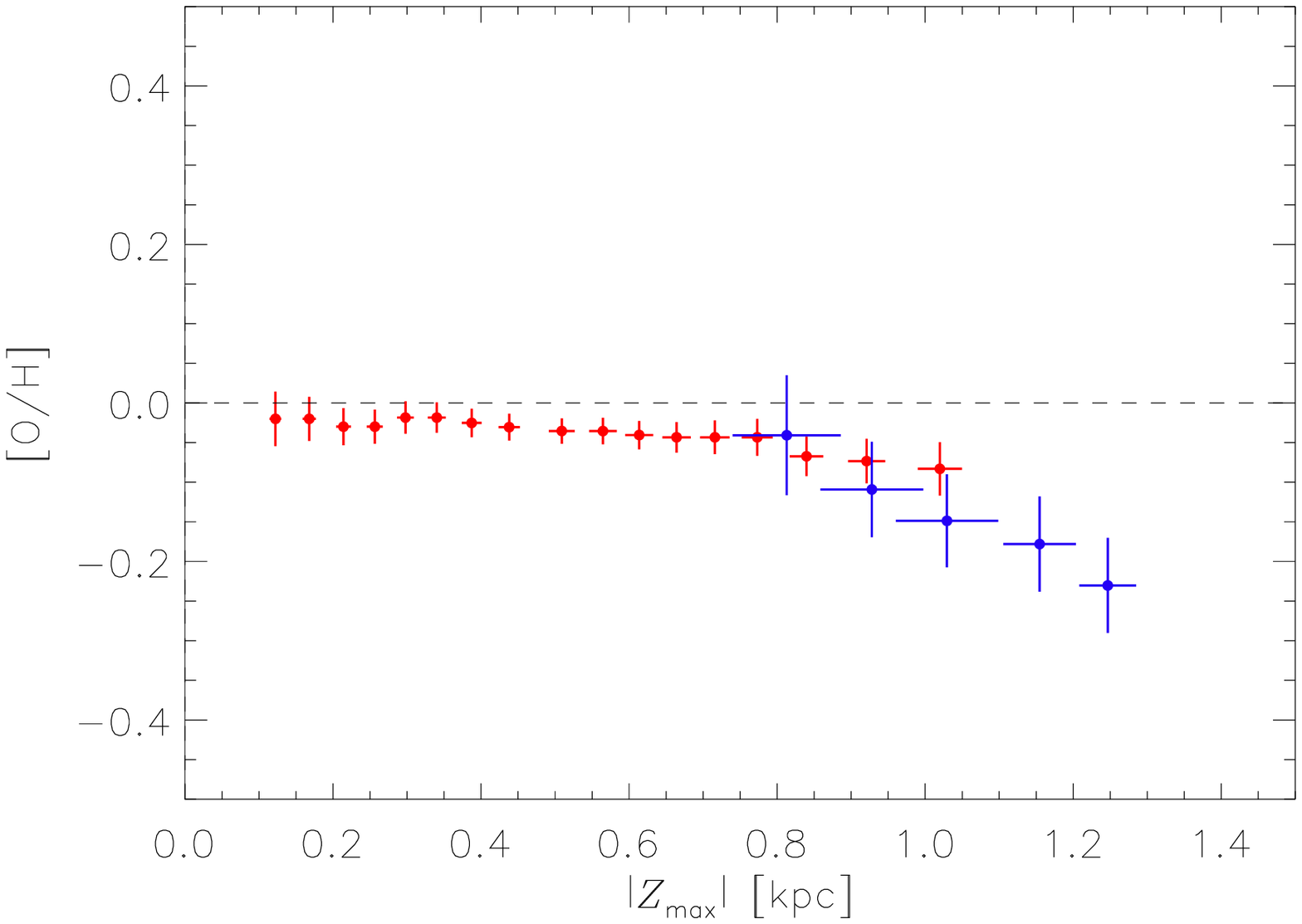} \\[-10pt]
\hspace*{-0.5cm}\includegraphics[width=0.45\textwidth]{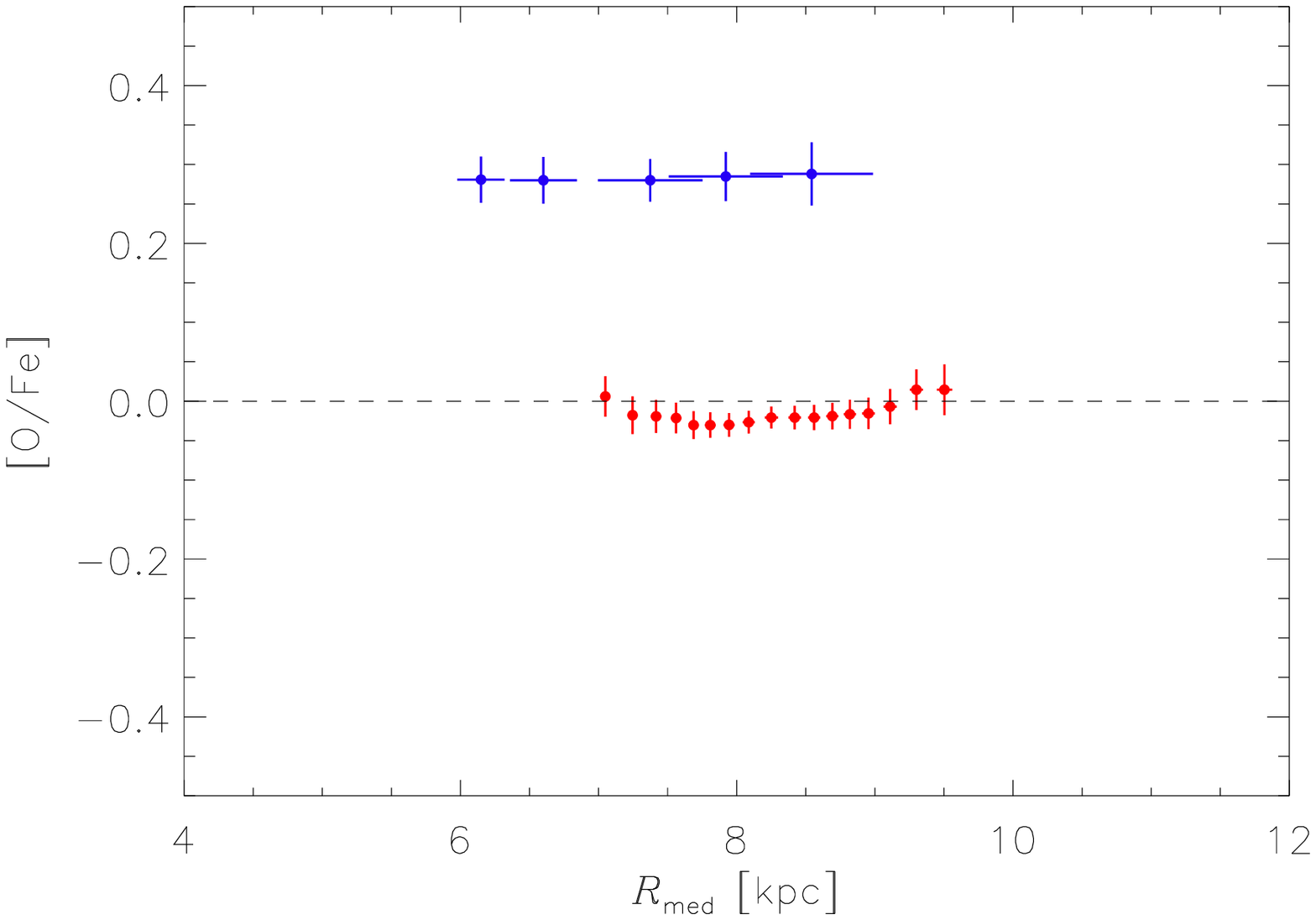} &
\hspace*{-0.5cm}\includegraphics[width=0.45\textwidth]{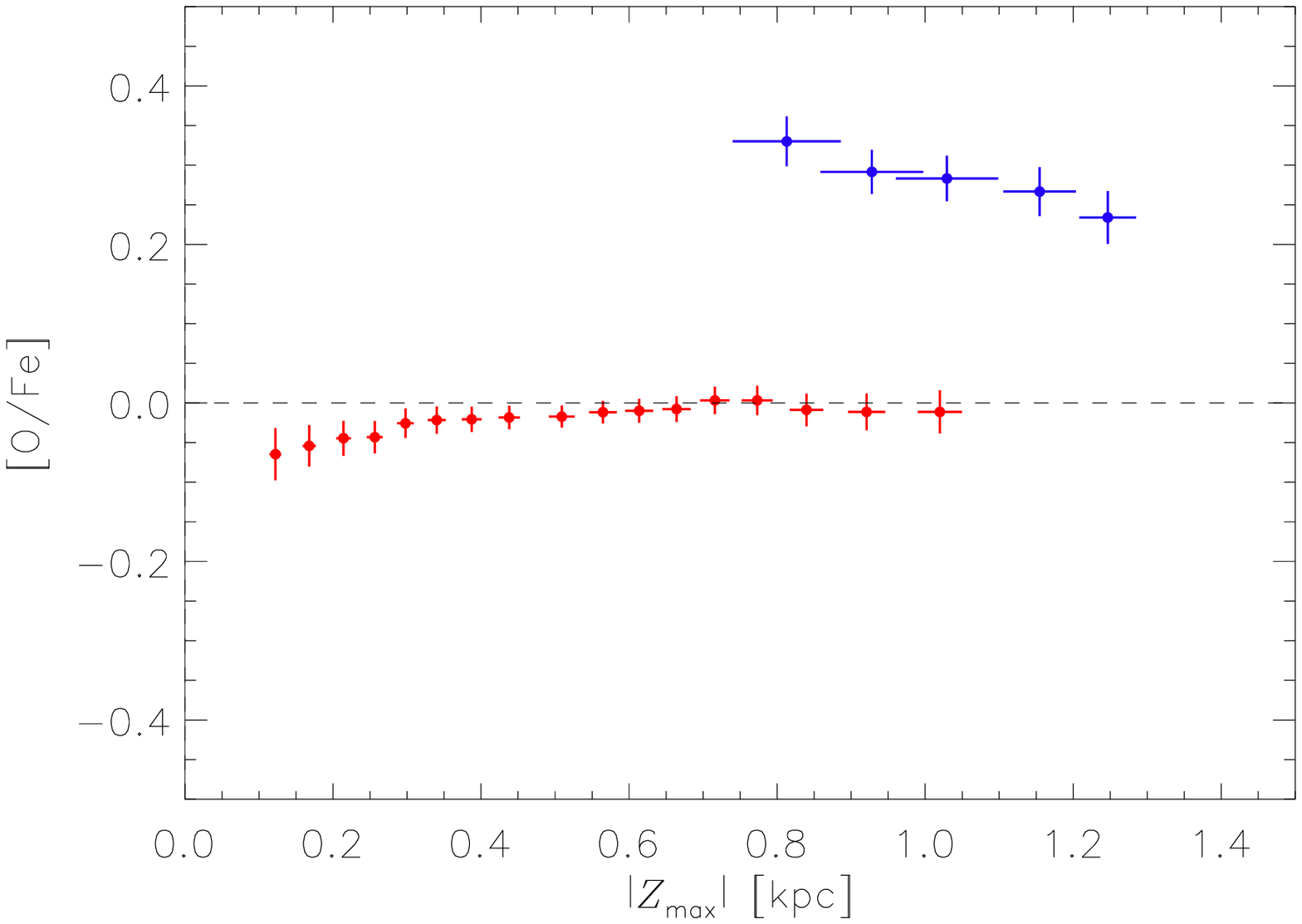} \\[-10pt] 
\hspace*{-0.5cm}\includegraphics[width=0.45\textwidth]{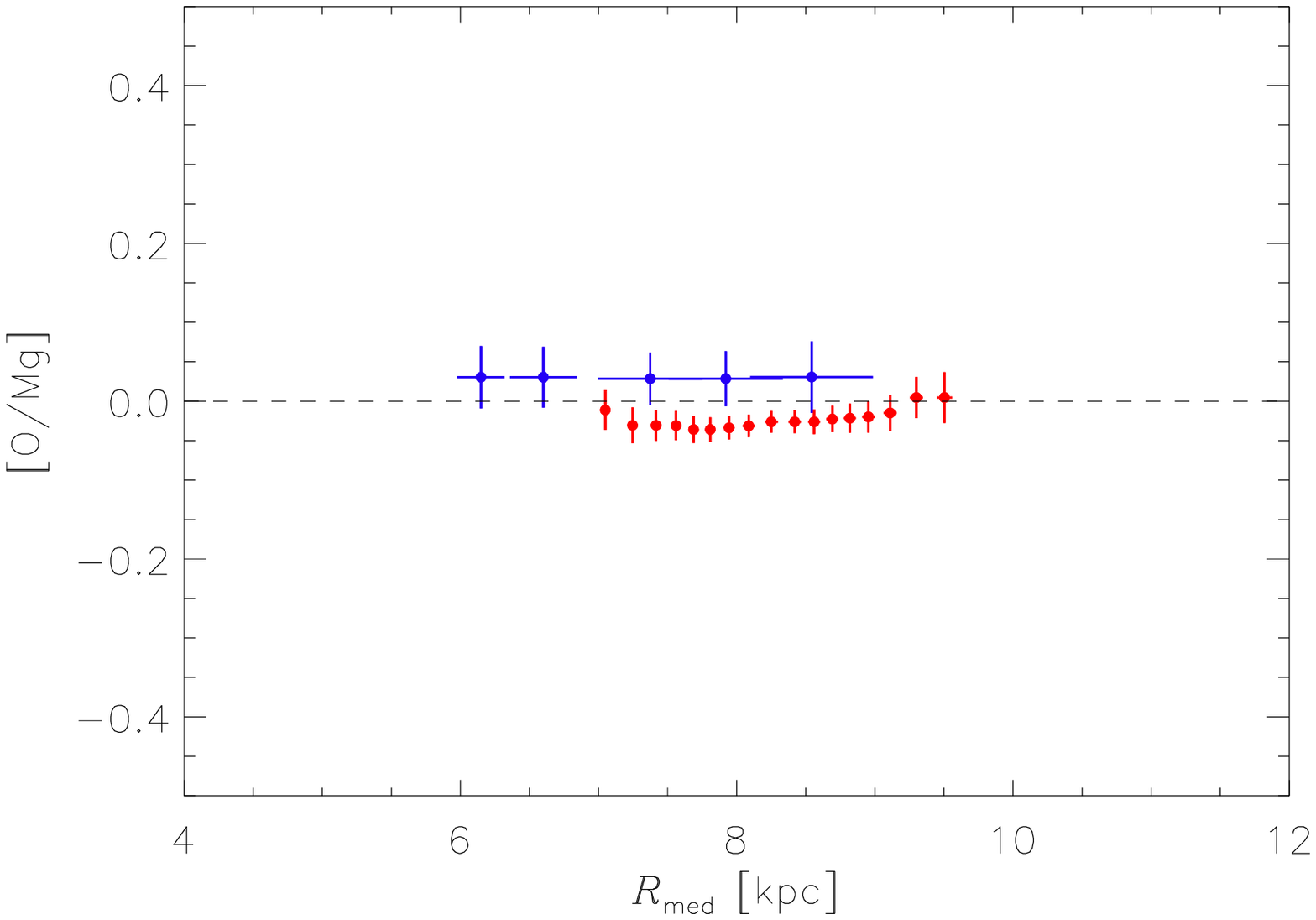} &
\hspace*{-0.5cm}\includegraphics[width=0.45\textwidth]{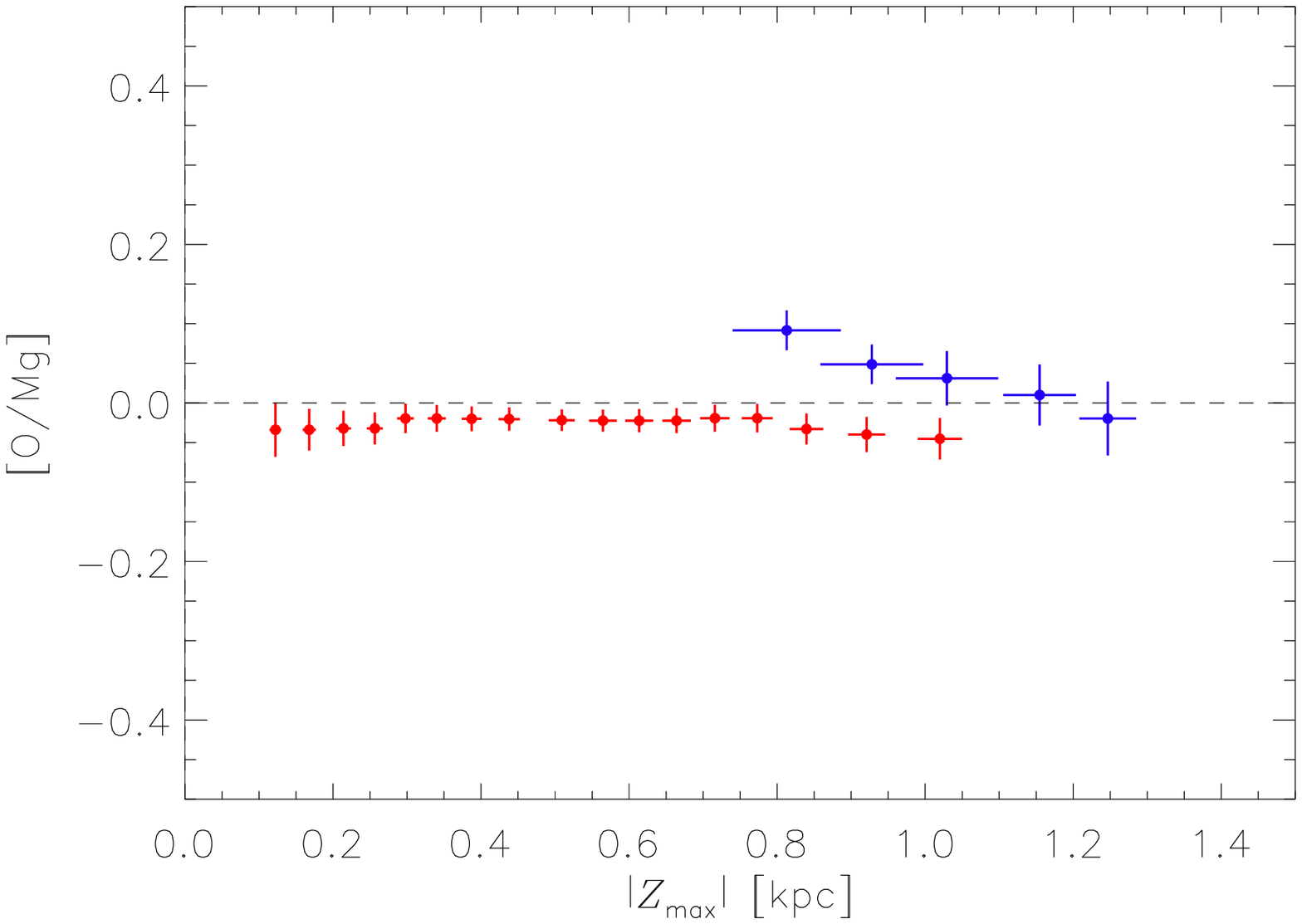} \\[-10pt] 
\hspace*{-0.5cm}\includegraphics[width=0.45\textwidth]{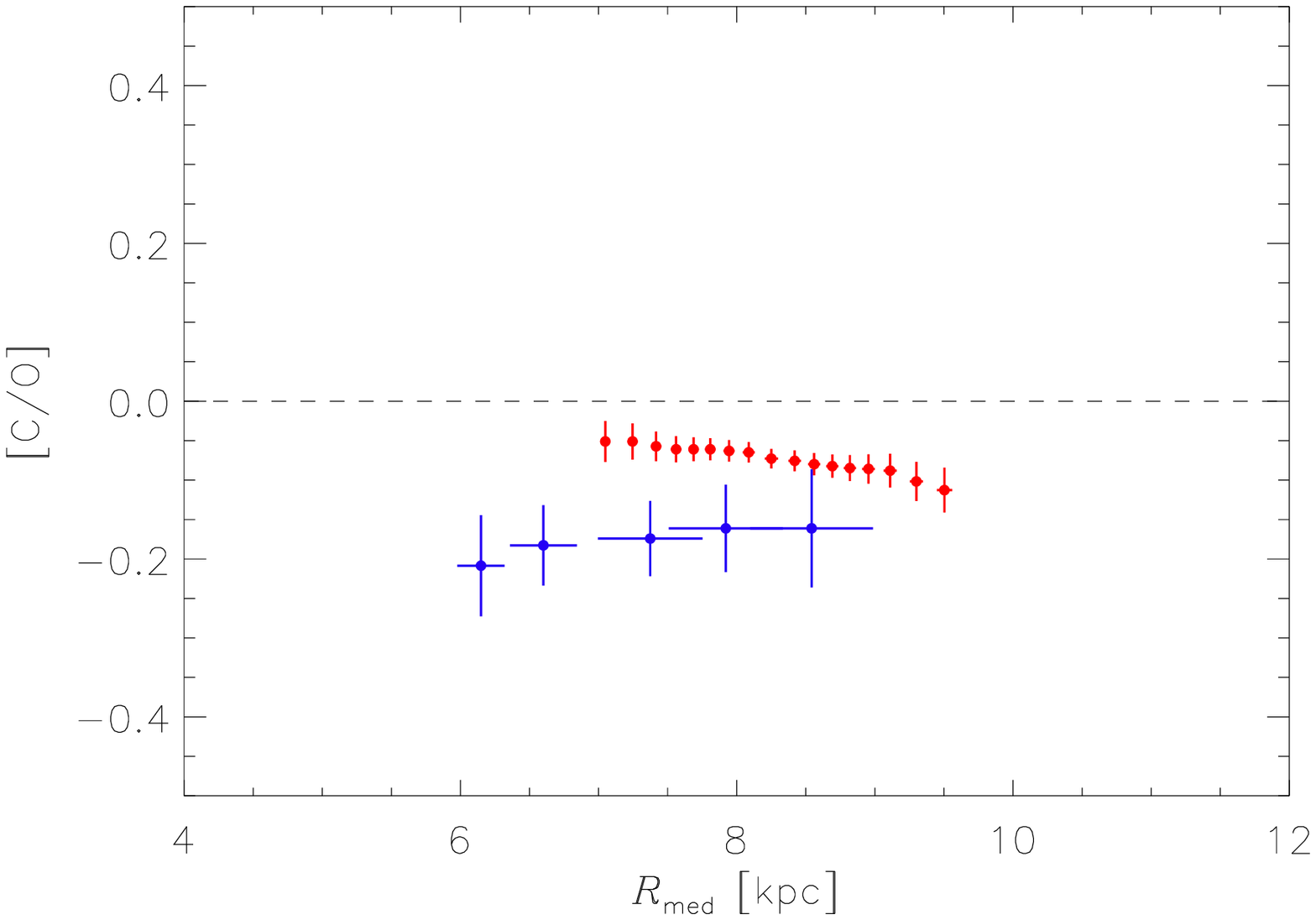} &
\hspace*{-0.5cm}\includegraphics[width=0.45\textwidth]{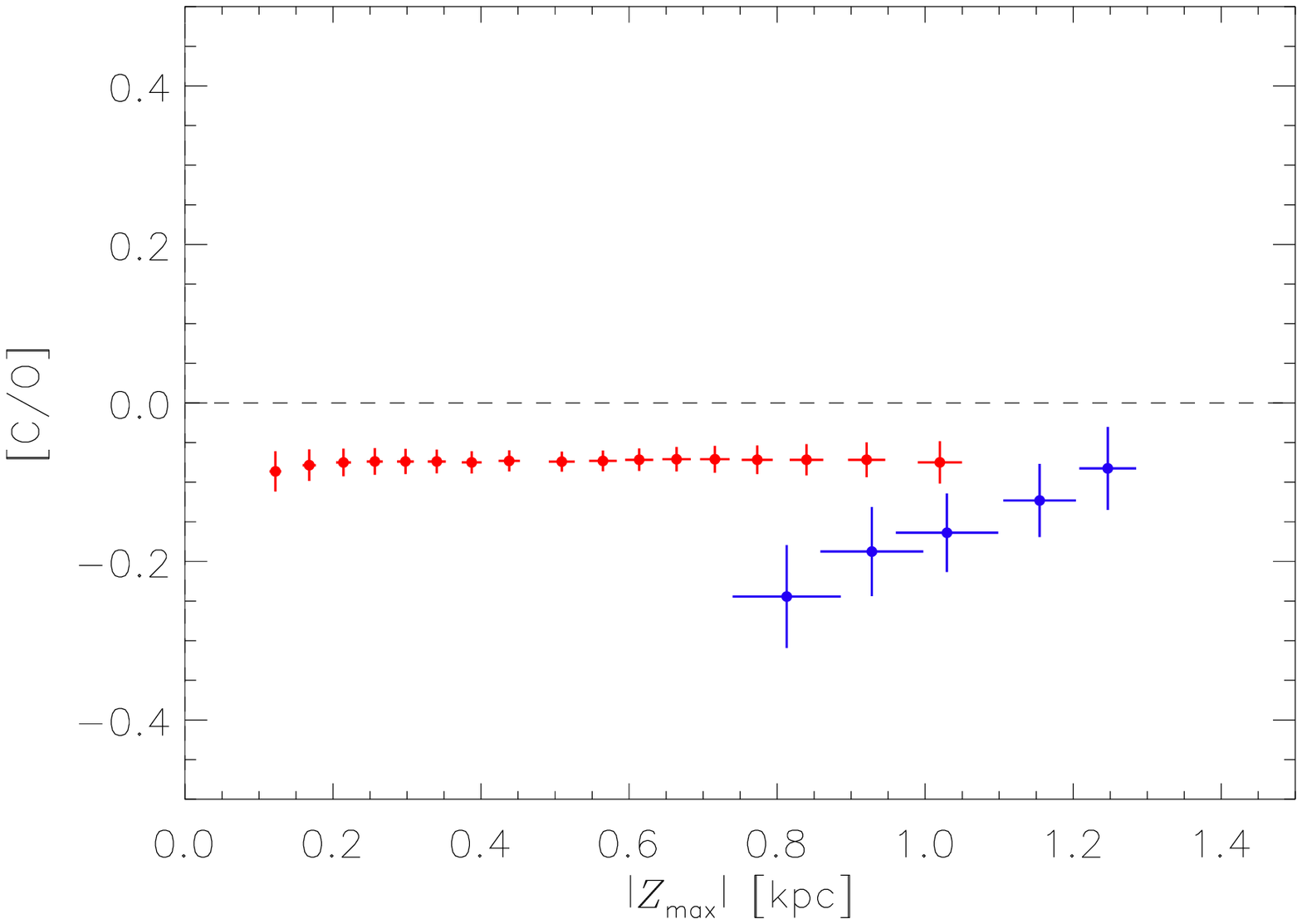} \\[-10pt] 

\end{tabular}
\caption{[O/H], [O/Fe], [O/Mg], and [C/O] (top to bottom)   versus $R_{\rm med}$  (left panels) and    $|Z_{\rm max}|$ (right panels) diagrams  for  thin (red) and thick (blue) disk star samples. 
\label{fig:OH_Rmed_Zmax}}
\end{center}
\end{figure}

Both observed radial and vertical stellar abundance distributions are particularly interesting 
tools to study the chemical enrichment history of the Galactic disks and their formation processes. Actually, the radial and vertical thin  and thick disk star trends collect
a lot of very important information to disentangle  the embedded aspects of inflow, star formation, outflow, and radial and vertical mixing that gave rise to the present state of the Milky Way disk (for example, a radial gradient of the disk that is produced  in an inside-out formation scenario 
can be partially reduced/cancelled by a radial mixing process).
The radial metallicity gradients  of the Milky Way disk have
been measured using a number of different stellar (e.g. 
Cepheids, open clusters) and interstellar medium tracers (HII regions, etc) which  represent the composition of the interstellar gas at the time 
they were formed \citep [e.g.][and references therein]{MIK14,CHE12} or at the present time, respectively, thus probing
the metallicity gradient at different epochs. In addition, the vertical gradients of the global metallicity and of different elements have also been derived by using different input databases \citep[see, for example][and references therein]{SCL14,BOE14,MAG17}. One of the causes of different gradient values, both radial and vertical, reported in the literature might be  attributed to different input data. In fact, gradients may be affected by the chemical evolution  of the Galaxy
 (i.e. by the time the analyzed objects were formed), by local inhomogeneities, moving groups and stellar streams.
 Moreover, the measured gradients may be biased by systematics and/or inhomogeneities in the spatial and age distributions of the input data samples, i.e. derived vertical gradients may be more affected by stars located in the outer or the inner galactic radius, $R$, and/or by younger or older stars if the $Z$ bins do not contain evenly distributed objects in $R$ and age. Therefore,
the interpretation of the derived gradients as result of the star formation history of the galactic disk(s) is very challenging.
In the following we will take advantage of the fact that the stars in our thin and thick disk stellar samples display, to some extent, different dependency on  $R_{\rm med}$ and  $|Z_{\rm max}|$ and span different age ranges (see Figures\,\ref{fig:OH_Rmed_Zmax}\,and\,\ref{fig:histo_Age}), but it is worthwhile to remark that our interpretation of results  may still be affected by systematics, in particular, for the thick disk stellar sample.

In Figure \,\ref{fig:OH_Rmed_Zmax}  we show the radial and vertical gradients of [O/H], [O/Fe], [O/Mg] and [C/O]  for the  thin and thick disk star samples using again  running average bins as in Figures\,\ref{fig:OFe_Mg}\,and\,\ref{fig:CO}. 

As far as the  trends as a function of  $R_{\rm med}$ are concerned,  we recall that FR20
do not have in their Galactic
potential deviations from axisymmetry, like those introduced by
the bar and the spiral arms, and do not take into account radial stellar migration. However,  $R_{\rm med}$ values can  provide an indication of the stellar birthplaces \citep{EDV93, ROC04} which could be  used, at least,  in a differential form.

As can be seen, the thin and thick disk stars fall in different  $|Z_{\rm max}|$ ranges: most of the thin disk bins  have $|Z_{\rm max}|<0.8$\,kpc while all the 
thick disk bins  have $|Z_{\rm max}|>0.8$\,kpc. Moreover, as discussed in the next Section,  most of the thin disk star bins  have  ages $< 10$\,Gyr while all the thick disk objects have ages $> 10$\,Gyr (see Figure\,\ref{fig:Age_Abu}).
Therefore, the trends with  $R_{\rm med}$ of our thin and thick disk stars may give hints on the radial gradients of relatively young stars at low  $|Z_{\rm max}|$ and of old stars at higher  $|Z_{\rm max}|$, respectively.

\begin{deluxetable}{l|cc|cc|cc|cc}
\tablecaption{Radial gradients \label{tab:gradR}}
\tablecolumns{10}
\tablewidth{0pt}
\tablehead{
\multicolumn{1}{c}{  } &
\multicolumn{1}{|c}{   $\frac{{\rm d[O/H]}}{{\rm dR_{med}}} $} &
\multicolumn{1}{c}{   $\chi_{\rm r}$} &
\multicolumn{1}{|c}{  $\frac{{\rm d[O/Fe]}}{{\rm dR_{med}}} $} &
\multicolumn{1}{c}{  $\chi_{\rm r}$} &
\multicolumn{1}{|c}{  $\frac{{\rm d[O/Mg]}}{{\rm dR_{med}}} $} &
\multicolumn{1}{c}{   $\chi_{\rm r}$} &
\multicolumn{1}{|c}{  $\frac{{\rm d[C/O]}}{{\rm dR_{med}}} $ } &
\multicolumn{1}{c}{   $\chi_{\rm r}$} \\
\multicolumn{1}{c}{  } &
\multicolumn{1}{|c}{  [dex/kpc] } &
\multicolumn{1}{c}{   } &
\multicolumn{1}{|c}{  [dex/kpc] } &
\multicolumn{1}{c}{   } &
\multicolumn{1}{|c}{  [dex/kpc] } &
\multicolumn{1}{c}{   } &
\multicolumn{1}{|c}{  [dex/kpc] } &
\multicolumn{1}{c}{  
}
}
\startdata
~~~~~Thin~~~~~ & -0.041$\pm$0.001 &  0.145 & 0.014$\pm$0.004 &   0.269  & 0.009$\pm$0.002 &   0.193  & -0.020$\pm$0.001 &   0.146 \\
~~~~~Thick~~~~~ & ~0.006$\pm$0.002 &  0.062 & 0.003$\pm$0.001 &  0.074 &  0.000$\pm$0.001 &   0.032 & 0.020$\pm$0.004 &   0.147 \\
\enddata
\end{deluxetable}

\begin{deluxetable}{l|cc|cc|cc|cc}
\tablecaption{Vertical gradients \label{tab:gradZ}}
\tablecolumns{10}
\tablewidth{0pt}
\tablehead{
\multicolumn{1}{c}{  } &
\multicolumn{1}{|c}{ $\frac{{\rm d[O/H]}}{{\rm d|Z_{max}|}}$} &
\multicolumn{1}{c}{  $\chi_{\rm r}$ } &
\multicolumn{1}{|c}{ $\frac{{\rm d[O/Fe]}}{{\rm d|Z_{max}|}}$ } &
\multicolumn{1}{c}{  $\chi_{\rm r}$} &
\multicolumn{1}{|c}{ $\frac{{\rm d[O/Mg]}}{{\rm d|Z_{max}|}}$} &
\multicolumn{1}{c}{  $\chi_{\rm r}$} &
\multicolumn{1}{|c}{ $\frac{{\rm d[C/O]}}{{\rm d|Z_{max}|}}$ } &
\multicolumn{1}{c}{   $\chi_{\rm r}$} \\
\multicolumn{1}{c}{  } &
\multicolumn{1}{|c}{ [dex/kpc] } &
\multicolumn{1}{c}{   } &
\multicolumn{1}{|c}{[dex/kpc] } &
\multicolumn{1}{c}{   } &
\multicolumn{1}{|c}{[dex/kpc] } &
\multicolumn{1}{c}{   } &
\multicolumn{1}{|c}{ [dex/kpc] } &
\multicolumn{1}{c}{  
}
}
\startdata
~~~~~Thin~~~~~ & -0.04$\pm$0.01 &   0.229 & 0.09$\pm$0.01 &  0.350 & 0.020$\pm$0.004 &  0.234 & 0.007$\pm$0.001 &  0.053 \\
~~~~~Thick~~~~~ & -0.41$\pm$0.03 &   0.203 & -0.20$\pm$0.02 &  0.310 & -0.24$\pm$0.02 &   0.278 & 0.35$\pm$0.02 &   0.141 \\
\enddata
\end{deluxetable}

By inspection of Figure\,\ref{fig:OH_Rmed_Zmax} we can observe that:
\begin{itemize}
    \item the [O/H] and [C/O] show negative gradients
     as a function of $R_{\rm med}$, 
    for the thin disk sample. The abundance ratios [O/H] and [C/O] decrease significantly with the mean Galactocentric distance, while the thick disk sample  exhibit  flat trends or perhaps very mild positive gradients as indicated by the fits of Table\,\ref{tab:gradR}. The positive radial gradient of [C/O] for the thick disk stars might not be real since it could actually be a result of the fact that our data bins correspond to groups of stars with different average $|Z_{\rm max}|$'s  and are actually reflecting the presence of the steep [C/O] vertical gradient shown in the bottom--right panel of Figure\,\ref{fig:OH_Rmed_Zmax}. 
   A dichotomy between thin and thick disk stars was found by several authors  
    \citep[e.g.][]{ALL06,KAT11,CHE12,BOE13,AND14,MIK14,REC14,KOR15,HAY15,HAY18}
    and could indicate that either the thick disk formed from a well mixed interstellar material or, if   gradients were  present, they
    disappeared because of a  migration effect more  efficient  in the thick disk than in the thin one
  \citep{MIN13}. \citet{EST05}, using recombination lines from
    echelle spectrophotometry  of 8 HII regions with Galactocentric distances in the range 6-10 kpc, derived for the [O/H] radial gradient a value  of  -0.044 $\pm$ 0.010\,dex/kpc. \citealt{RUD06},
    also based on  the analysis of HII regions,  derived a value of -0.060\,dex/kpc and of -0.041 dex/kpc by using optical and infrared data, respectively. More recently,
    \citet{WAN19}, used  spectra of 101 HII regions in the Galactic anti-center area from the Large Sky Area Multi-Object Fiber 
     Spectroscopic Telescope (LAMOST), and derived the oxygen abundance gradient with a slope of -0.036 $\pm$ 0.004\,dex/kpc.
   \citet{WEN19}, using the National Radio Astronomy Observatory Karl G. Jansky Very Large Array to observe
    $\sim$8-10 GHz hydrogen radio recombination line and radio-continuum emission toward 82 Galactic H II regions,  find an oxygen abundance
    gradient across the Milky Way disk with a slope of $-0.052 \pm 0.004$\,dex/kpc. All these values are in good agreement, within the uncertainties, with our slope value (the small errors given in 
    Table\,\ref{tab:gradR} are due to the fact that we fitted the abundances of the average bins and not of the individual stars). Of course, we compare our gradients with those from HII regions which, actually, reflect the present situation, while in our trends we mix stars of somewhat different ages with only a rough separation  between  young/thin and old/thick disk populations.
    As a further comparison \citet{CES07} found a gradient of -0.035\,dex/kpc for O for a Galactocentric distance from 4 to 14 kpc in good agreement with our results.
     
     A negative radial gradient of [C/O] has been observed by \citet{EST05} from HII region recombination lines. They found a steeper gradient, $\frac{{\rm d[C/O]}}{{\rm d|Z_{max}|}}=-0.058 \pm 0.018$ than us. On the theoretical side  \citet[][]{CAR05} show that, by varying carbon, nitrogen, and oxygen yields, Galactic chemical evolution models predict slopes between -0.005\,and\,-0.068\,dex/kpc.
     
      We notice that most of the [O/H] and all the [C/O] values of the thin disk bins are negative suggesting that solar neighbourhood stars have less oxygen and C/O than the Sun. To explain these results \citet{NIE12}
    state that the  Sun  may be an immigrant to its current Galactic neighbourhood, i.e. the birth place of the Sun  could be at an inner Galactocentric distance ($R=\sim$ 5–6\,kpc) where higher metallicity values were reached earlier in cosmic history. If this is true, they claimed that "A telltaling signature is left only in the C/O ratio" (but they  do not discuss [O/H]). Our results suggest that a solar [O/H] value should be characteristic of $R_{\rm med}=\sim 7$\,kpc while solar [C/O] would imply $R_{\rm med} \lesssim 6$\,kpc but we recall again that, due to the simplified Galactic potential used by FR20 to compute stellar orbits,  $R_{\rm med}$  gives only a rough indication of the actual stellar birthplaces;
    \item  The [O/Fe] and [O/Mg] versus  $R_{\rm med}$ show flat (or slightly positive for the thin disk star bins) trends (see Table\,\ref{tab:gradR}).  
     \citet{MIK14} find a slightly positive slope of [$\alpha$/M] in their clean thin disk sample (0.014 $\pm$ 0.004\,dex/kpc) and a flat trend  in their clean thick disk sample (-0.005 $\pm$ 0.006\,dex/kpc). Unfortunately, their results cannot directly be compared with our [O/Fe] trends since their $\alpha$ elements do not include oxygen.
     The [O/Mg] versus  $R_{\rm med}$ show even flatter trends and a much smaller systematic difference between the values for thick and thin disk stars  than the [O/Fe]. These results might indicate that SNIa not only produce in the thin disk an enrichment of Fe but also a small increase of Mg while oxygen seems to be enriched only by CCSNe.

\end{itemize}

For the abundance trends as a function of  $|Z_{\rm max}|$ we identify the following features:
\begin{itemize}
\item  \rm{[O/H]} bins show a negative vertical gradient with slope increasing going from lower (thin disk star bins) to higher (thick disk star bins) $|Z_{\rm max}|$ (see Table\,\ref{tab:gradZ} where the thin disk fits refer to bins with $|Z_{\rm max}|<0.8$\,kpc). Similar behaviours for $\frac{{\rm d[Fe/H]}}{{\rm d|Z_{max}|}}$ and $\frac{{\rm d[Si/H]}}{{\rm d|Z_{max}|}}$ were found by \citet{BOE14}. In an analysis of $\alpha$-elements \citet{MIK14}, found $\frac{\rm d\alpha}{{\rm d|Z|}}=+0.036 \pm 0.006$\,dex/kpc for their thin disk clean sample. However, we want to remark that these authors  did not include oxygen in their $\alpha$ mixture;
\item  \rm{[O/Fe]} and [O/Mg] thin disk star bins show positive trends with $|Z_{\rm max}|$ (see Table\,\ref{tab:gradZ}) which could be explained by the increasing contribution of iron and magnesium  by SNIa (we recall that our stars at low $|Z_{\rm max}|$
are also the youngest ones as shown in Figure\,14 of FR20).
The thick disk star bins have negative trends like [O/H] but with lower slopes. The fact that  [O/Mg] shows a smaller difference between the average values of thick and thin disk star bins than [O/Fe] (as already seen in the radial gradient panels) is a consequence of the greater yields of iron than of magnesium of SNIa;
\item  \rm{[C/O]} show an almost flat trend for the thin disk star bins suggesting that both carbon and oxygen were produced by massive stars. The positive trend of the thick disk star bins reflects the steep negative trend of [O/H] with  $|Z_{\rm max}|$ shown in the upper-right panel\footnote{The flat behaviour of [C/H] versus  $|Z_{\rm max}|$ for the same objects, as  checked by the authors, does not affect the [C/O] trend.}.
\end{itemize}

\subsection{{\rm [O/H], [O/Fe],[O/Mg]}, and {\rm [C/O]} trends with age}
\label{sec:age}

\begin{figure}[htbp]
\includegraphics[width=0.95\textwidth]{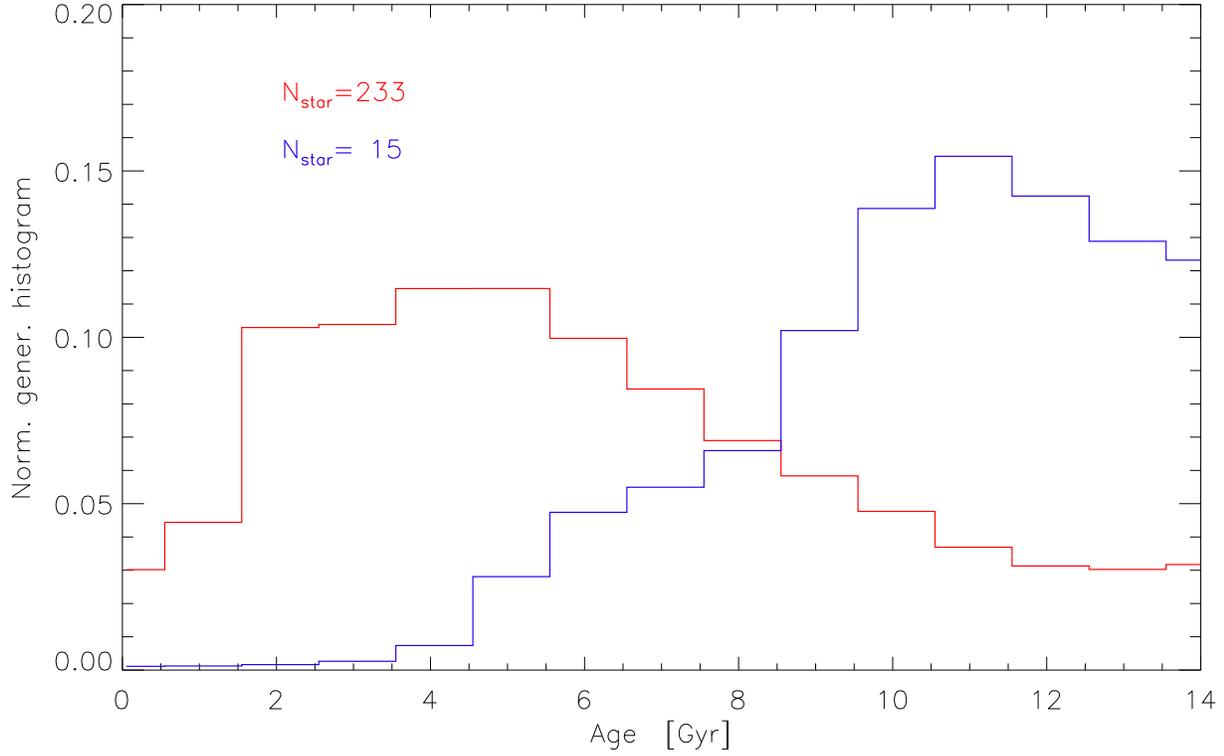}
\caption{Normalized generalized Age histograms   for  thin (red) and thick (blue) disk star samples. 
\label{fig:histo_Age}}
\end{figure}

FR20 give nominal ages  for   233  and 15 stars belonging to our thin or thick disk star sample, respectively. 
The thick disk stars are, in general,  older than members of the thin disk.
Figure\,\ref{fig:histo_Age} shows the normalized generalized
age histograms of both populations, built by summing the individual age probability distributions,  for the thin (red) and thick (blue) disk stars. The extended wings and therefore, the overlap of the two distributions, are likely due to the large uncertainties affecting
individual stellar ages. 
Most of the thin disk stars span a range from 2 to 8\,Gyr while the thick disk stars are mostly older than 8\,Gyr. This is in agreement with the common understanding of Galactic chemical evolution based on serial and two-infall models \citep[e.g.][]{CHI97, ROM10, GRI17, CAL09} and   with chemical abundance studies, matched with asteroseismologic age determinations,
which suggest a delay of $\sim 4$\,Gyr between the first and second accretion episodes which gave origin to the MW
disks \citep{HAYW15,SNA15,SPI19}.

\begin{figure}[ht]
\begin{center}
\begin{tabular}{cc}
\hspace*{-0.5cm}\includegraphics[width=0.5\textwidth]{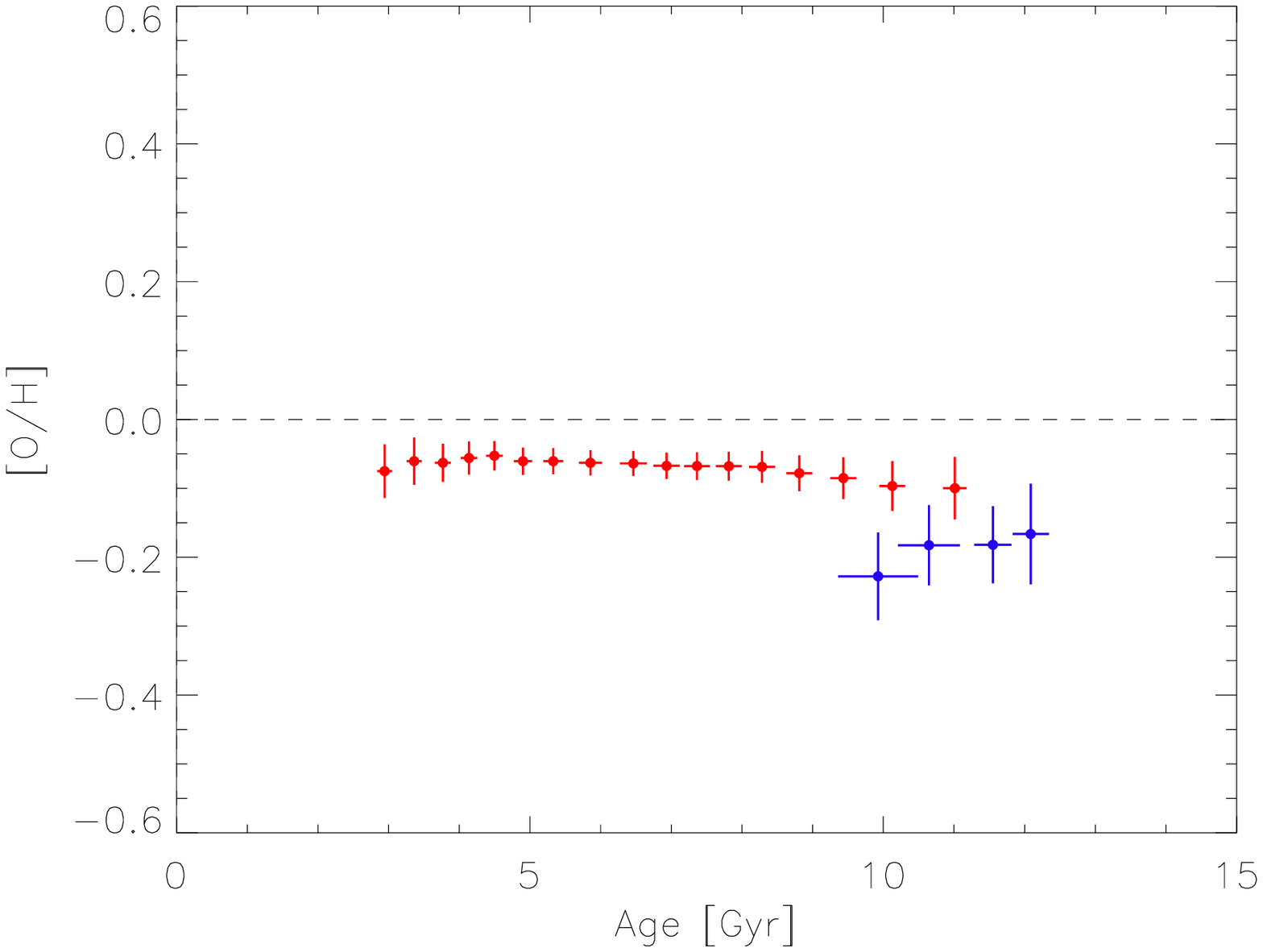}  &
\hspace*{-0.5cm}\includegraphics[width=0.5\textwidth]{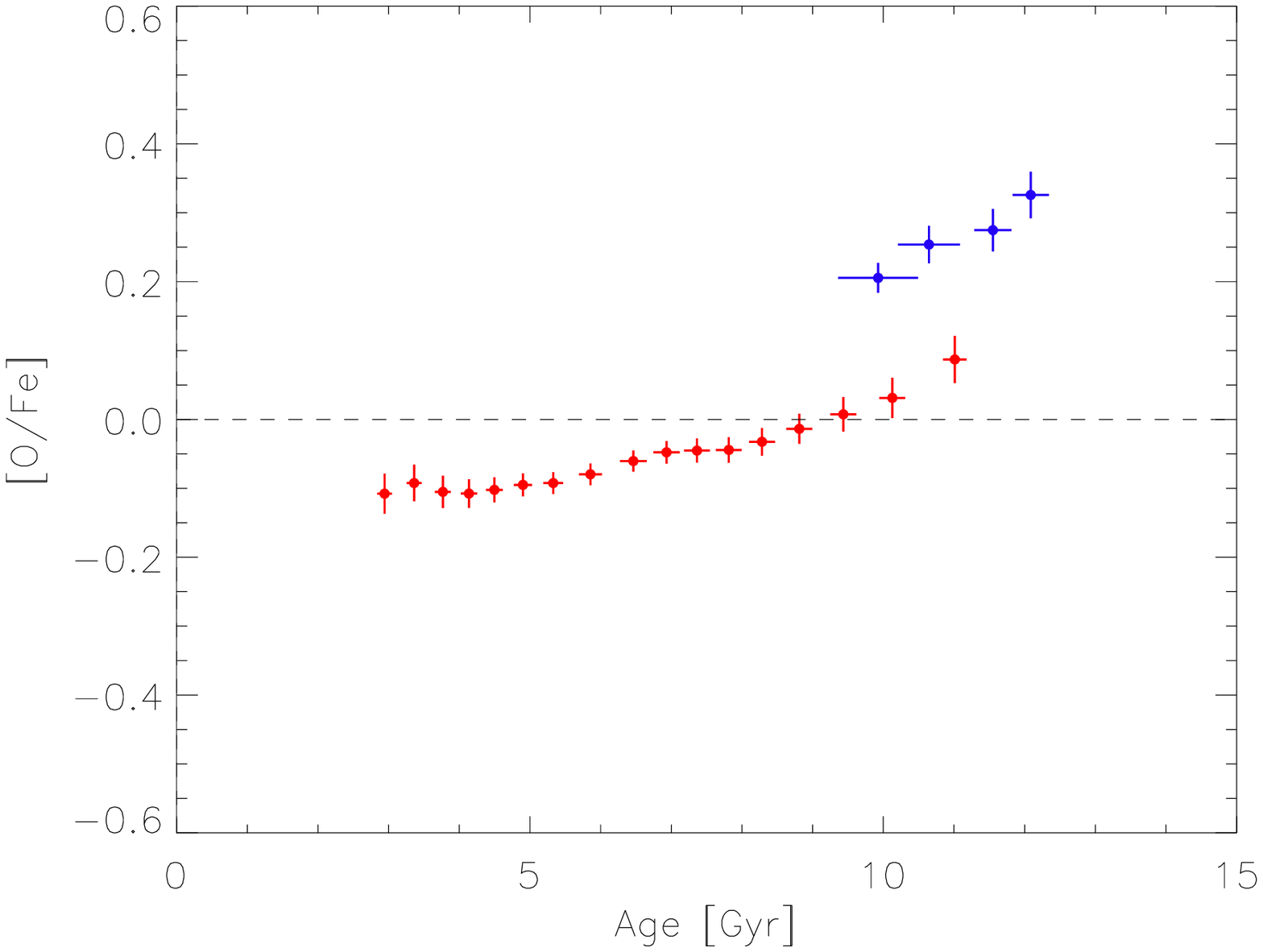} 
\\[-10pt]
\hspace*{-0.5cm}\includegraphics[width=0.5\textwidth]{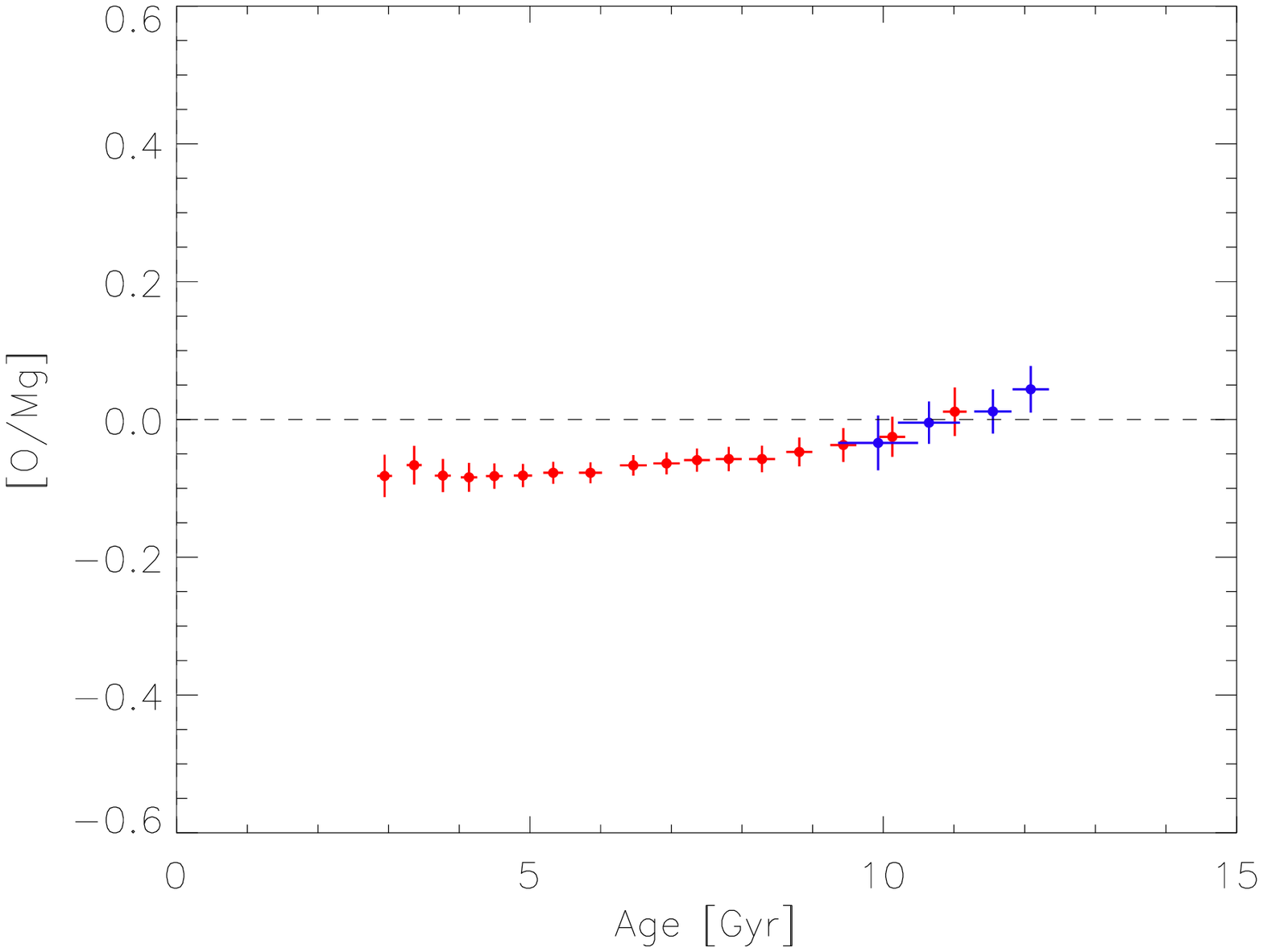}  &
\hspace*{-0.5cm}\includegraphics[width=0.5\textwidth]{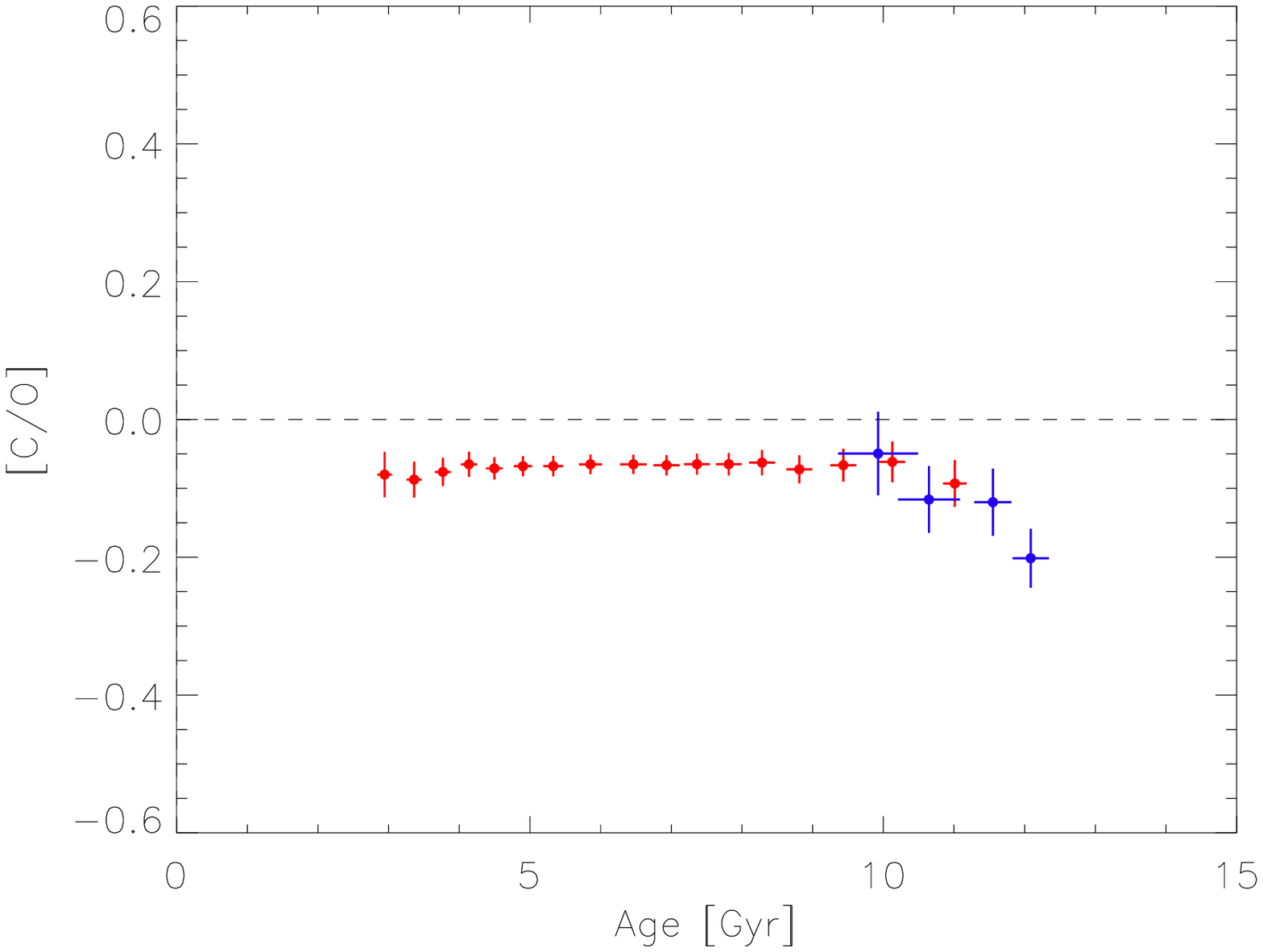}  
\end{tabular}
\caption{[O/H]  (upper-left panel),  [O/Fe] (upper-right panel),  [O/Mg] (lower-left panel),
and [C/O] (lower-right panel) vs Age diagrams for  thin (red) and thick (blue) disk samples.
\label{fig:Age_Abu}}
\end{center}
\end{figure}

Figure\,\ref{fig:Age_Abu}  shows the trends of [O/H], [O/Fe], [O/Mg] and [C/O] versus age. As for previous figures, we plot in red and blue the thin and thick disk stellar samples, with
the bins constructed using a running average.

In Figure\,\ref{fig:Age_Abu} we identify the following properties of abundance ratios versus age:

\begin{itemize}
    \item a  difference in the age intervals spanned by thin (in general younger) and thick (older) disk star bins is clearly evident  with only a small overlap between  10\,and\,11\,Gyr. Recalling the intrinsic uncertainties on FR20 age estimates it is possible that such an overlap could actually be less pronounced;
   \item both thin and thick disk stars show almost flat trends of [O/H] versus age. 
     A similar behaviour is observed by \citet{DEL19} that, by using the oxygen abundances from \citet{BER15}, derived for the HARPS-GTO sample two similar flat trends of [O/H] with age for thin and thick disk stars.
     The small negative slope ($-0.004 \pm 0.001$) for our thin disk star bins and the slightly larger positive one ($0.028 \pm 0.004$) for our thick disk star bins with age are likely to be  remnant of the radial gradient and of the vertical gradient (plus selection effects) shown in Figure\,\ref{fig:OH_Rmed_Zmax} (red points in the upper-left panel and blue points in the upper right panel, respectively). The higher average value for the thin disk stars indicates that the thin disk formation started only after many CCSNe's had already produced O-enrichment of the interstellar medium;  
    
    \item both thin and thick disk stars show increasing trends of [O/Fe] and [O/Mg] with age with an almost zeroed slope for the youngest objects. Since our stars do not show a strong spatial gradient in  [O/Fe] and [O/Mg] (see the central panels of Figure\,\ref{fig:OH_Rmed_Zmax}), we think that they  really indicate abundance ratio trends with age. There is a jump from higher to lower [O/Fe] values in the common age region while the [O/Mg] sequences merge smoothly. 
    The steeper trends at oldest ages confirm that oxygen is indeed mainly produced at the early phases of Galaxy formation by  massive stars. Such correlations were also  in other  [$\alpha$/Fe] ratios \citep[see for example][and references therein]{NIS18,DEL19}. 
    In fact, steep age trends  at old ages and flat trends at young ages should be expected for [O/Fe] if the CCSNe's are the predominant (or even the only) suppliers of oxygen in the interstellar medium while a less clear behaviour can be foreseen  in the case of  other $\alpha$-elements, which are also partially released by SNIa  \citep[e.g][]{CAR05,NOM13}. We also found steeper slopes for [O/Fe] than for [O/Mg] which is consistent with the fact that SNIa release much more iron than magnesium in the interstellar medium.   The slopes of [O/Fe] and [O/Mg] versus age for our thin disk star bins are 0.019 and 0.01\,dex/Gyr, respectively, and are in qualitative agreement with the results by \citet{DEL19}  who found  0.026 and 0.02\,dex/Gyr using an HARPS-GTO sample.
    The smooth merge of the thin and thick trends of [O/Mg] seems to indicate that  [O/Mg] should be preferred to [O/Fe] as a proxy of age and  its shape suggests that the correlation is non-linear in the whole age interval from 3 to 12\,Gyr;
   
      \item  the [C/O] trends show a steep decrease  for the oldest objects (age $\gtrsim 10$\,Gyr) and an almost flat trend for younger objects.  As in the case of [O/H] this behaviour with age can be actually the result of hidden spatial [C/O] gradients (see the bottom panels of Figure\,\ref{fig:OH_Rmed_Zmax}).
    This interpretation agrees with the results by \citet{NIS15}, who, using HARPS spectra of 21 solar twins in the solar neighborhood,  showed that the C/O ratio evolves very little with time,  although [C/Fe] and [O/Fe] ratios  evolve significantly. In that paper it is shown a good agreement  between the data (see his Figure\,11) and  the predictions of the Galactic chemical evolution model by \citet{GAI15}, in which the dominant contribution of C and O comes from massive stars. It is also argued that C/O has only a moderate rise during the first $\sim$5 Gyr due to an increasing contribution of carbon from stars of lower mass. This is also in agreement with what was found by FR20. 

   \end{itemize}

\section{Comparison with other surveys}
\label{sec:surveys}
In this section we compare some of our results presented in  Sections\,\ref{sec:OFe} and \ref{sec:CO}  
with the median trends  derived for other large stellar samples of the solar neighbourhood, namely, APOGEE \citep[][hereafter, WE19]{WEI19}, GALAH DR2 (second data release) \citep[][hereafter, GR19]{GRI19}), and one constructed upon HIRES Keck data \citep[][hereafter, BR16]{BRE16}.  WE19, who adopted Mg rather than Fe as their reference element,  separated their stars, on the basis of their  [Fe/Mg],  into two populations (15 and 996 low-[Fe/Mg] or high-$\alpha$ and high-[Fe/Mg] or low-$\alpha$, respectively, with [O/Mg] determinations), and derived the median trends of [X/Mg] versus [Mg/H] in each population. 
Likewise, GR19  found 220 and 1146, respectively, of  high-$\alpha$ and low-$\alpha$ stars (renamed by the authors 
low-Ia and high-Ia stars) with trustworthy carbon and oxygen abundances, and  derived  median sequences of [X/Mg] versus [Mg/H].
 We adopted, the same kind of approach
in order to separate the stars in the BR16's sample. Actually, we used  the chemical criterion described
in FR20, i.e. we used, in the same range of $T_{\rm eff}$ and $\log g$ of our GES stars, the positions of the BR16 stars in the [Mg/Fe]-[Fe/H] plane to obtain two samples of 460 and 22 low- and high-$\alpha$ stars, respectively. In the following we will consider the WE19, GR19 and BR16 high-$\alpha$ (low-Ia) and the low-$\alpha$ (high-Ia) stars as  the analogous of our thin  and thick disk sample, respectively.

\begin{figure}[htbp]
\includegraphics[width=0.5\textwidth]{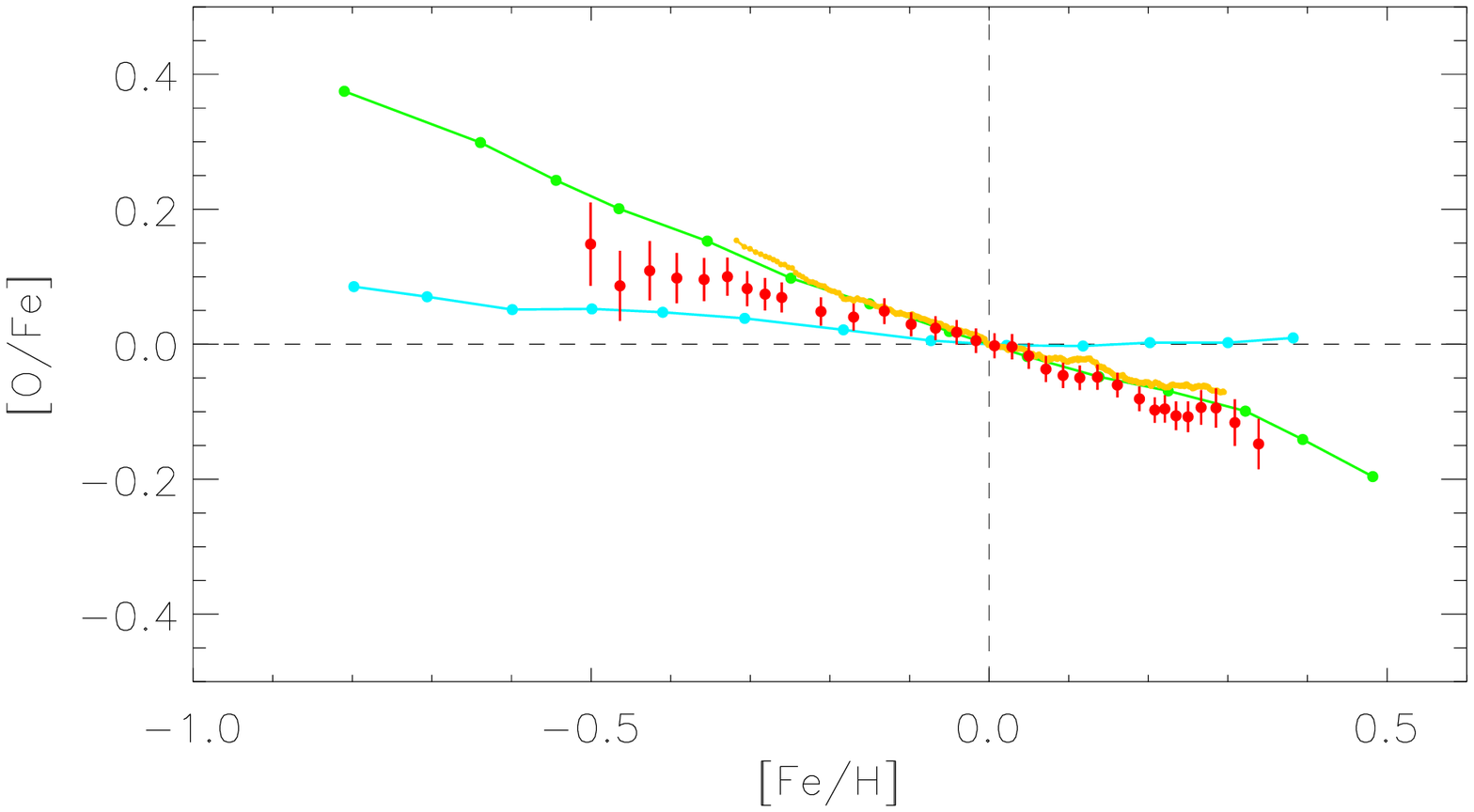}
\includegraphics[width=0.5\textwidth]{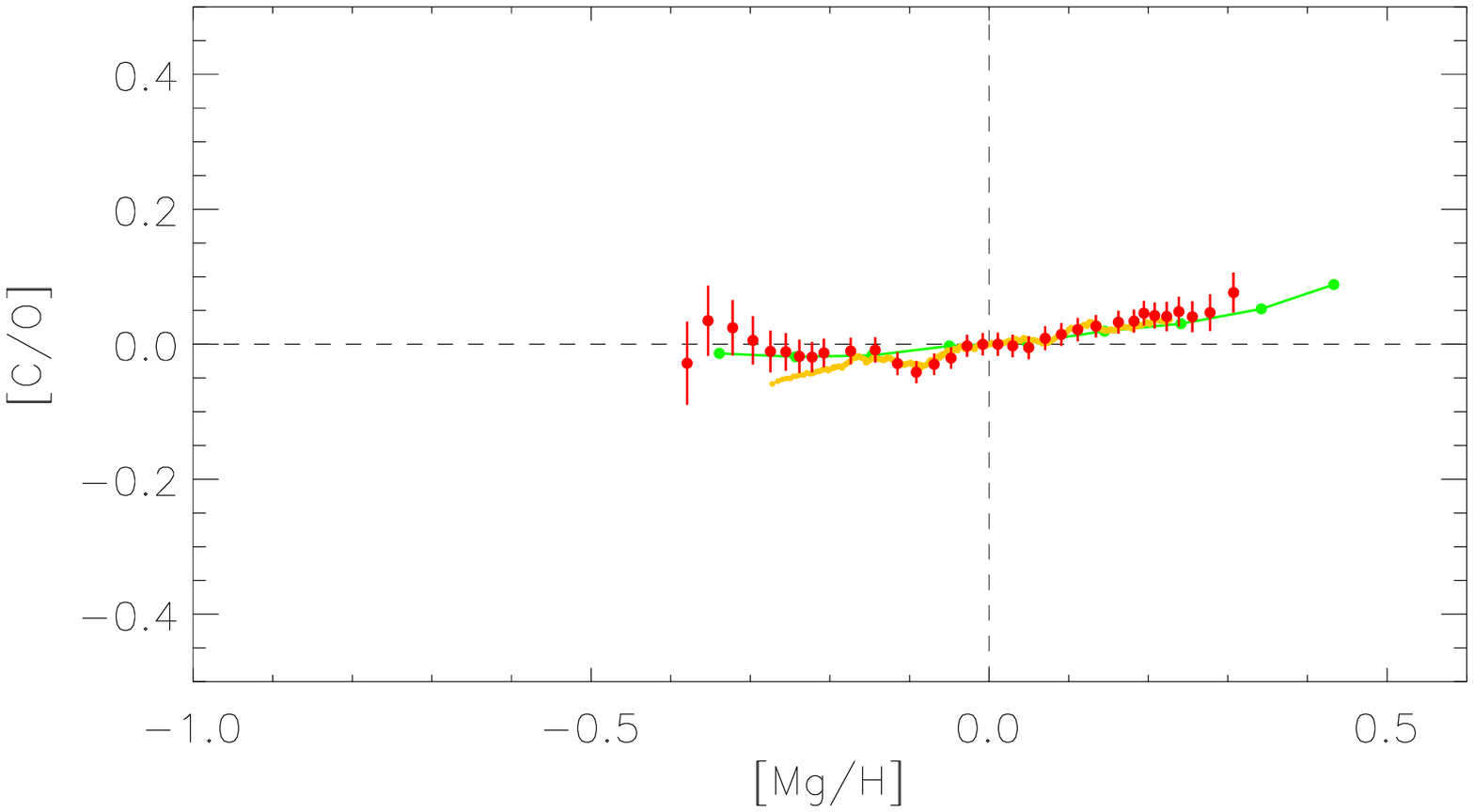}
\includegraphics[width=0.5\textwidth]{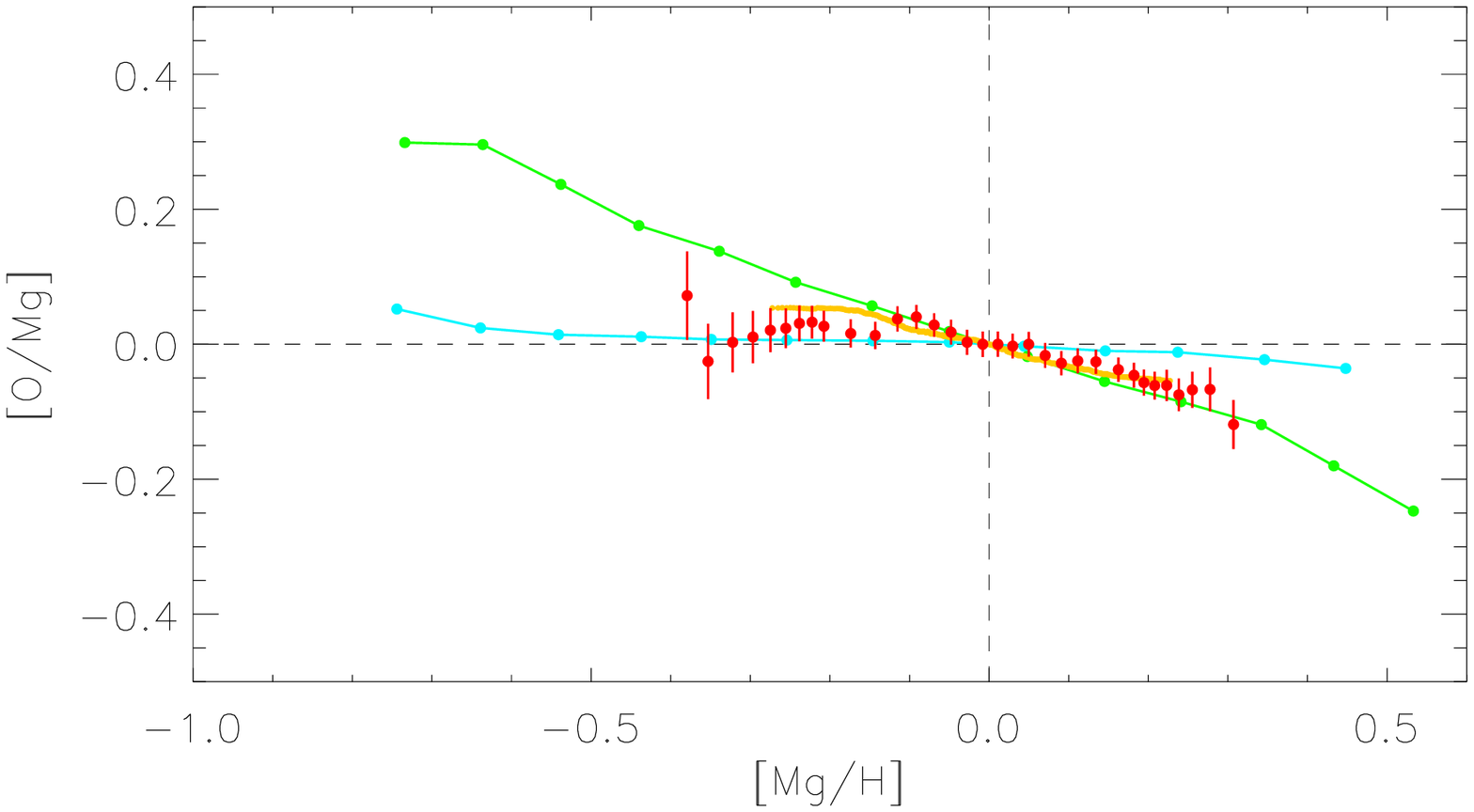}
\includegraphics[width=0.5\textwidth]{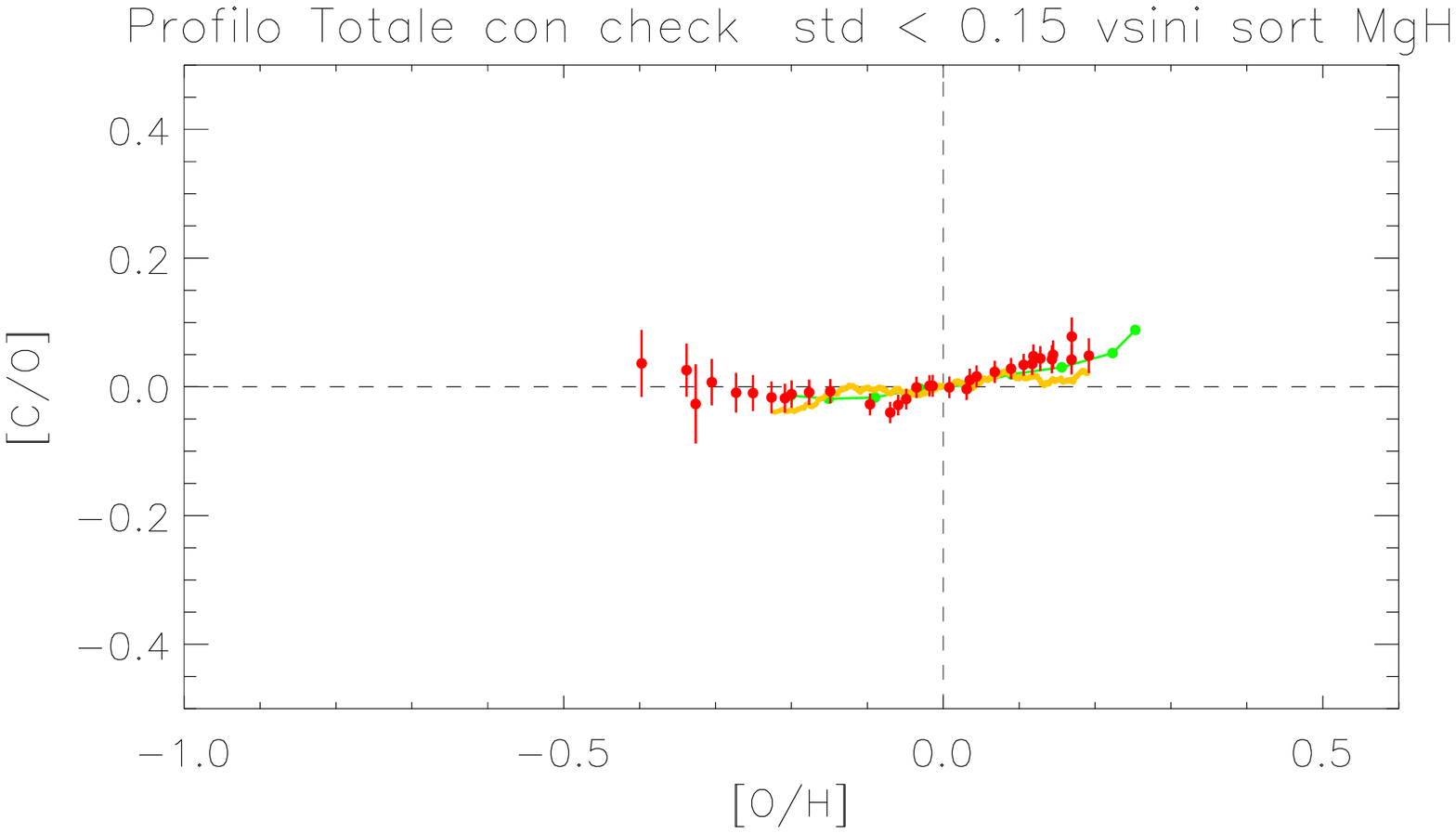}
\caption{Comparison between our  [O/Fe]-[Fe/H]  (top-left panel),   [O/Mg]-[Mg/H]
(bottom-left panel), [C/O]-[Mg/H] (top-right panel), and [C/O]-[O/H] (bottom-right panel) trends and WE19, GR19, and BR19 ones. Red symbols  represent our  thin disk star bins,  as in previous figures. Continuous  lines represent trends of  thin disk stars from  WE19 (light blue), GR19 (green), and BR16  (yellow) data. In the two plots of [C/O] abundance ratios, WE19 data are not available.
\label{fig:trend_comp}}.
\end{figure}

In Figure\,\ref{fig:trend_comp} we compare our [O/Fe]-[Fe/H],   [O/Mg]-[Mg/H],  [C/O]-[Mg/H] and  [C/O]-[O/H] trends with those from WE19, GR19, and BR19. Because of the small stellar sample of thick disk stars, we focus on the thin disk members. As in previous figures, red points symbols  are used to identify  our  thin disk star bins while continuous  lines represent   the trends from  WE19 (light blue), GR19 (green), and BR16  (yellow). GR19 applied zero-point off-sets to
their GALAH DR2 data points and trends so that stars with [Fe/Mg] = [Mg/H]$\simeq$ 0 also have [X/Mg]$\simeq$ 0.
To make the comparison easier, we adopted a similar approach and computed  also for our, WE19, and BR16 data the zero-point off-sets in order to have [O/Fe] and [O/Mg]$\simeq 0$
for thin disk (low $\alpha$) stars with [Fe/H] or [Mg/H]$\simeq$ 0. The data illustrated in  Figure\,\ref{fig:trend_comp} are plotted after applying the proper zero-offset. 

From the two left panels  we can clearly see  that  the WE19 trends of [O/Fe] and [O/Mg] differ 
significantly from the others. WE19 [O/Fe] versus [Fe/H] and [O/Mg] versus [Mg/H] 
trends are almost flat  while the GR19 and the BR16  trends  are both 
sloped  and nearly superimposed for 
${\rm [Fe/H]}\gtrsim -0.3$\,dex and
for ${\rm [Mg/H]} \gtrsim -0.1$\,dex, respectively. Our data agree with GR19 and 
BR16 results for  ${\rm [Fe/H]} \gtrsim -0.15$\,dex and for ${\rm [Mg/H]} \gtrsim -0.1$\,dex , but fall between the GR19 and WE19 [O/Fe] trends for lower [Fe/H] and between  BR16  and WE19 [O/Mg] trends for lower [Mg/H] eventually merging with WE19 for ${\rm [Mg/H]} \gtrsim -0.35$\,dex.

In the right panels of  Figure\,\ref{fig:trend_comp} we compare our [C/O] versus [Mg/H] and [O/H] thin disk star bins with the corresponding trends by GR19 and BR16\footnote{APOGEE also measures C but WE19 do not present carbon results since their stars are mainly giants where original carbon surface abundances are altered by mixing  with internal processed material during the 
first dredge-up phase \citep[see][]{IBE65}, therefore no WE19 trends are present in the two right panels.}. In both diagrams there is a very good agreement of our data with the GR19 and BR16 trends which do also very well agree with each other. 

Several reasons can explain the observed differences between the  trends shown in Figure\,\ref{fig:trend_comp}. For instance, different  thresholds in the [Mg/Fe]-[Fe/H] plane adopted by GR19 and WE19 to separate the thin disk stars from the thick disk stellar bins with respect to those used by us,  may led to slightly higher mean values of the [Mg/H] ratio at a given [Fe/H] value. Moreover, we recall that our two samples were selected not using only a chemical criterion. Therefore, the already  emphasized  well-known problem of the different definition of the Galactic thin and disk may have introduced inhomogeneity in the trends we are comparing. It is also important to bear in mind that stellar abundances depend in different ways on the galactic position and age of the stars. WE19 stated that their [X/H] trends with  [Mg/H] do not vary with the different Galactic positions, it cannot be excluded that the differences in the WE19, GR19, BR16 trends and in our data can be due to the different mixture of radial and vertical positions, and of ages in the studied samples.

\section{Summary and Conclusions}
\label{sec:conclusions}

We have measured oxygen abundances of 516 FGK dwarf stars in the Galactic disk(s).  The stellar sample was extracted from the iDR5 Gaia-ESO Survey and  constitute a sub-sample of the FGK  dwarf set  whose carbon abundances were derived in FR20. In this paper, we  derived the oxygen abundances by using the spectral synthesis of the forbidden oxygen line at 6300\,\AA~  taking special care to account for the contribution to the observed
blend of the two isotopic lines of NiI at 6300.336\,\AA. 
Only those stars with spectra unaffected by telluric lines and/or by  bad removal of the sky emission line in the spectral region were considered. The abundance determinations ([O/Fe]) were conducted through a comparison of  the observed spectra with ``on-the-fly''  synthetic spectra computed with the SPECTRUM code from fully consistent atmosphere structures obtained by using the ATLAS12 program.

In order to discriminate between the thin  and thick disk objects we used   three different selection approaches based on (a) the stellar positions in the [Mg/Fe]-[Fe/H] plane, (b) the stellar Galactic velocities, and (c) orbital parameters. Then, we merged the results of these three methodologies and, after discarding the few stars with discordant classifications, we obtained two samples of 376 and 20 stars classified,  by at least one of the three methods, as thin  or thick disk stars, respectively. 
In this paper we compared the stars in the two samples by deriving trends  of [O/H],  [O/Fe], [O/Mg], and [C/O] versus [Fe/H] and [Mg/H] and, for the stars with Galactic orbits and ages computed in FR20, also versus the stellar position, in and above the Galactic plane, and age. 
Our main results can be summarized as follows:
\begin{itemize}
   \item[-]  our thin and thick disk stars span different age intervals, with the thick disk members being older than those of the thin disk as already found in many previous studies; 
   
   \item[-] the  [O/H] values versus [Fe/H] show a large scatter with reduced $\sqrt(\chi^2)$ of the linear fits $\sim 3$ suggesting that a significant part of the observed scatter may be astrophysical;
   
   \item[-] similarly to other $\alpha$-elements, oxygen abundance ratios ([O/H],  [O/Fe] versus [Fe/H]) show systematic differences between   thin  and thick disk stars: ({\it i}) thick disk stars have larger O abundance than the thin disk stars at the same [Fe/H] in the  range covered by our data; ({\it ii}) thin disk stars show a clear monotonic decreasing trend of [O/Fe] with increasing metallicity, even at the metal-rich regime, in agreement with previous results \citep[e.g.][]{CAS97,CHE03,BEN04,ECU06,BEN14,AMA19} and with Galactic chemical evolution models \citep[e.g.][]{CHI03}.
   This last result suggests that the oxygen enrichment of the interstellar medium is produced only by CCSNe with no  evidence of  contributions from SN\,Ia or AGB stars that would cause a flattening of [O/Fe] at [Fe/H] $\simeq$ 0 like those  observed in other $\alpha$-elements  \citep{BEN04,BEN14};
   
   \item[-] at  [Mg/H]$>0.0$ the thin disk stars show a decrease of [O/Mg] with [Mg/H] indicating that  oxygen and magnesium  do not evolve in lockstep, hence these chemical species should have different origins, i.e.  magnesium  should be released, to some extent, also by SN\,Ia and/or AGB stars while oxygen is only enriched by CCSNe;
   
 \item[-] the spatial and age trends cannot be easily disentangled in particular for the thick disk objects, due to the paucity of available data. Our thin and thick disk samples, due to the difference in age of the component stars, can  also be interpreted as two age groups, with a boundary at $\sim$ 9\,Gyr. The young group clearly follows negative radial gradients in [O/H] and in [C/O] and  positive vertical trends for [O/Fe] and [O/Mg]. The second group shows almost flat radial trends  and negative vertical gradients for [O/H], [O/Fe] and [O/Mg]; the positive trends found for [C/O], as mentioned in Section\,\ref{sec:orbits}, should be taken with caution. [O/Fe] and [O/Mg] show positive trends with age for both the thin and the thick disk stellar samples, with a smooth merge of the two trends for [O/Mg], suggesting that this abundance ratio can be a good proxy of stellar ages. The [O/H] trend of the thin disk stars is almost flat and suggests that the thin disk formed after most of the enrichment of oxygen in the interstellar medium by CCSNe already occurred. The positive trends of [O/H] and [C/O] with age for the thick disk sample need further investigation (more objects) to clarify whether they are real or the result of a combination of statistical effects and of the observed vertical gradient of [O/H].
  
    \item[-] the thin disk star trend of [C/O] versus [Fe/H] shows a negative value of  [C/O] at [Fe/H]$\simeq$0 which implies that the Sun has a higher C/O value than the solar neighbourhood. We adopt a solar ratio of 0.54 \citep{GRE07} while our thin disk sample indicates a lower C/O value of $\sim$0.45.
   \citet{NIE12}, by using  early-type B stars infer, for the mean present-day chemical composition of the cosmic matter in the solar neighbourhood,   C/O=0.37. They attribute this result, which is in qualitative agreement with our, to  an outward migration of the Sun  in the galactic disk and to the effects of Galactic chemical evolution  and abundance gradients. The C/O ratio, therefore, reveals imprints left from the Sun’s original chemical composition.
 \end{itemize}

 We can conclude that our results confirm the hypothesis that CCSNe are the principal or, possible, the only accountable source of enrichment of oxygen in the Galaxy disks and that [O/Mg] can be a better indicator of stellar age than [O/Fe].
   
The data presented in this paper, including the derived abundances and atmospheric parameter values, are part of the full data--set from  GES. The spectra used here are a subsample of the over 200,000 reduced spectra of more than 100,000 stars observed and processed by the Gaia-ESO teams and are available through the ESO archive. The astrophysical parameters and abundances will be  available shortly.

\acknowledgments
This work is based on data products from observations made with ESO Telescopes at the La Silla Paranal Observatory under programme ID 188.B-3002. 
These data products have been processed by the Cambridge Astronomy Survey Unit (CASU) at the Institute of Astronomy, 
University of Cambridge, and by the FLAMES/UVES reduction team at INAF/Osservatorio Astrofisico di Arcetri. 
These data have been obtained from the Gaia-ESO Survey Data Archive, prepared and hosted by the Wide Field Astronomy Unit, Institute for Astronomy, 
University of Edinburgh, which is funded by the UK Science and Technology Facilities Council.
This work was partly supported by the European Union FP7 programme through ERC grant number 320360 and 
by the Leverhulme Trust through grant RPG-2012-541. We acknowledge the support from INAF and Ministero dell' Istruzione, 
dell' Universit\`a e della Ricerca (MIUR) in the form of the grant ``Premiale VLT 2012''. The results presented here benefit from 
discussions held during the Gaia-ESO workshops and conferences supported by the ESF (European Science Foundation) through the GREAT Research Network Programme.

This work has made use of data from the European Space Agency (ESA) mission Gaia (\url{https://www.cosmos.esa.int/gaia}), processed by the Gaia
Data Processing and Analysis Consortium (DPAC, \url{https://www.cosmos.esa.int/web/gaia/dpac/consortium}). Funding for the DPAC
has been provided by national institutions, in particular the institutions participating in the Gaia Multilateral Agreement.

This work received partial financial support
from PRIN MIUR 2010--2011 project ``The Chemical and dynamical Evolution of the Milky Way
and Local Group Galaxies'', prot. 2010LY5N2T and  by the National Institute for Astrophysics (INAF) through the grant PRIN-2014 (``The Gaia-ESO Survey''). 

M.C. thanks financial support from CONACyT grant CB-2015-256961. V.A. is supported by FCT - Funda\c{c}\~ao para a Ci\^encia e Tecnologia (FCT) through national funds and by FEDER through COMPETE2020 - Programa Operacional Competitividade e Internacionaliza\c{c}\~ao by these grants: UID/FIS/04434/2019; UIDB/04434/2020; UIDP/04434/2020; PTDC/FIS-AST/32113/2017 \& POCI-01-0145-FEDER-032113; PTDC/FIS-AST/28953/2017 \& POCI-01-0145-FEDER-028953. V.A. also acknowledge the support from FCT through Investigador FCT contract nr.  IF/00650/2015/CP1273/CT0001.  T.B. acknowledges financial support by grant No. 2018-04857 from the Swedish Research Council. U.H. acknowledges support from the Swedish National Space Agency (SNSA/Rymdstyrelsen).

This research uses the facilities of the Italian Center for Astronomical Archive (IA2) operated by INAF.

%

\vspace{5mm}
\facilities{VLT:Kueyen, UVES}


\software{SPECTRUM  \citep[v2.76f;][]{GRA94}, 
          ATLAS12 \citep{KU05a}
          }




\bibliography{mybiblio}{}
\bibliographystyle{aasjournal}



\end{document}